%% file: Iceland.tex
\documentclass[a4latex]{article}
\usepackage[paper=a4paper,left=20mm,right=20mm,top=20mm,bottom=20mm]{geometry}
\usepackage{amssymb,latexsym,amsfonts,url,rotating,dcolumn,ifthen}
\usepackage[fleqn]{amsmath}
\usepackage{pslatex}
\usepackage{epstopdf}
\usepackage[mathscr]{eucal}
\usepackage[usenames]{color}
\usepackage[dvips]{hyperref}
\usepackage[T1]{fontenc}
\usepackage{longtable}
\usepackage{graphicx}
\usepackage{epsfig}
\usepackage{fancyhdr}
\usepackage{setspace}
\usepackage{geometry}
\usepackage{subfig}
\usepackage{multicol}
\usepackage{libertine}
\usepackage[absolute]{textpos}      
\usepackage[document]{ragged2e}
\begin{document}
\input{titlepage_van.tex}
\setlength{\parindent}{15pt}
\section*{Abstract}
In 1979 following a decade of hyperinflation, Iceland introduced
a form of long term lending known as Ver{\dh}trygg{\dh} l\'{a}n, 
negatively amortised, index-linked loans whose outstanding principal is 
increased by the rate of the consumer price inflation index(CPI). These loans 
subsequently became the primary form of long term lending within Iceland by 
commercial banks and other institutions. 

The loans were part of a general government policy which used indexation 
to the CPI to address the economic consequences of the hyperinflation. 
This progressively linked increases in prices, wages and eventually loans 
directly to the CPI.  Although most other forms of indexation were subsequently 
removed, loan indexation has remained, and these loans now comprise the 
majority of mortgages in Iceland. Although it is still often argued that 
index-linked loans helped to stop the hyperinflation, these arguments are
typically based on high level macro-economic interpretations of the Icelandic
economy, they fail the scientific test of providing specific mechanisms to 
support their claims.  In this paper we take the opposite approach, and present a detailed 
analysis of the monetary mechanics used for the loans, and their effect on 
the Icelandic economy, based on a complete model of their interaction 
within the banking system at the fundamental level of all its 
transactions - the double entry bookkeeping level. 

Based on this analysis there appears to be no evidence or mechanism that 
would support the claim that index-linked loans reduce or stop inflation. 
Instead
our research shows that the bookkeeping treatment of these loans within
the banking system directly contributes to the banking system's monetary 
expansion rate, and hence index-linked loans act to \emph{increase} the 
inflation rate to which they are linked, rather than reducing it. They 
consequently create a positive feedback loop within the banking 
system's monetary regulation operating directly
on the money supply.  Our results indicate that the additional 
monetary expansion caused by these loans has ranged between 4\% to 12\% per 
anum over the past 20 years, and continuing excessive inflation can be expected
until these loans are removed from the economy. As a consequence
borrowers with these loans will find eventual repayment difficult 
if not impossible.  Since the feedback into  monetary expansion only 
occurs at annual CPI rates above approximately 2\%, we suggest one
solution would be to stabilise the money supply to 0\% growth, and we 
explore some ways this could be achieved 
by modifying the Basel Regulatory Framework within the Icelandic Banking 
System.

\section*{Introduction}
When macro-economic comparisons are made between countries, it is 
rarely the case that differences in the
types of lending, particularly long term lending, are discussed.
There are significant differences though.  In the United States mortgage 
lending is predominantly made at a fixed rate for the duration of the loan, 
while in the United Kingdom
long term lending rates are linked to the Bank of England's rate,
and can vary during the loan\footnote{Keynesian policy measures that
operate directly on the centrally set interest rate consequently have 
far more immediate effect in their country of origin, than in countries
with fixed rate regimes.}. In Iceland, as a result of of measures 
to combat a decade long hyperinflation that followed the 
collapse of the Bretton Wood's accord in the 1970's, the majority of long term 
lending between 1979-2013 has used an 
unusual form of financial instrument, the index-linked loan 
or Ver{\dh}trygg{\dh} l\'{a}n. These are loans 
structured with a base fixed interest rate, and an additional component 
which directly links their outstanding principal to the Consumer Price 
Inflation index (CPI)\footnote{See Appendix A for the calculation 
formula for these loans.}. 
\par
Indexed linked loans were the predominant form of long term consumer 
and commercial
lending in Iceland between 1980-2008 and are still available today, 
although they are now competing with variable and fixed rate loans. 
They are typically structured as 25 or 40 year loans, with a fixed
interest rate, and an additional component that increases 
the outstanding principal of the loan based
on the CPI.  This latter is structured to negatively amortise 
over the first half of the loan.  As is typical with negatively
amortised loans, while the initial repayments
are lower than comparable fixed or variable rate loans, the total cost 
of the loan is considerably higher due to the growth of the outstanding 
principal during the negatively amortised years of the loan.
\par
Considerable confusion has surrounded the technical aspects of the loans,
with particular respect to the calculation of the indexation component,
and the associated indexation index which
has been modified several times.  Historical information on the interest 
and indexation rates applied to these loans has proved difficult to find, 
and we draw the reader's attention to the qualification we must put on the 
sources used in this paper. In all cases where alternatives exist, we have 
chosen the most conservative series - and consequently the examples given 
in this paper may be under-estimates of the actual situation of the debt load
imposed by these loans.
\par
\begin{figure}[ht]
\begin{center}
\includegraphics[width=10cm, clip]{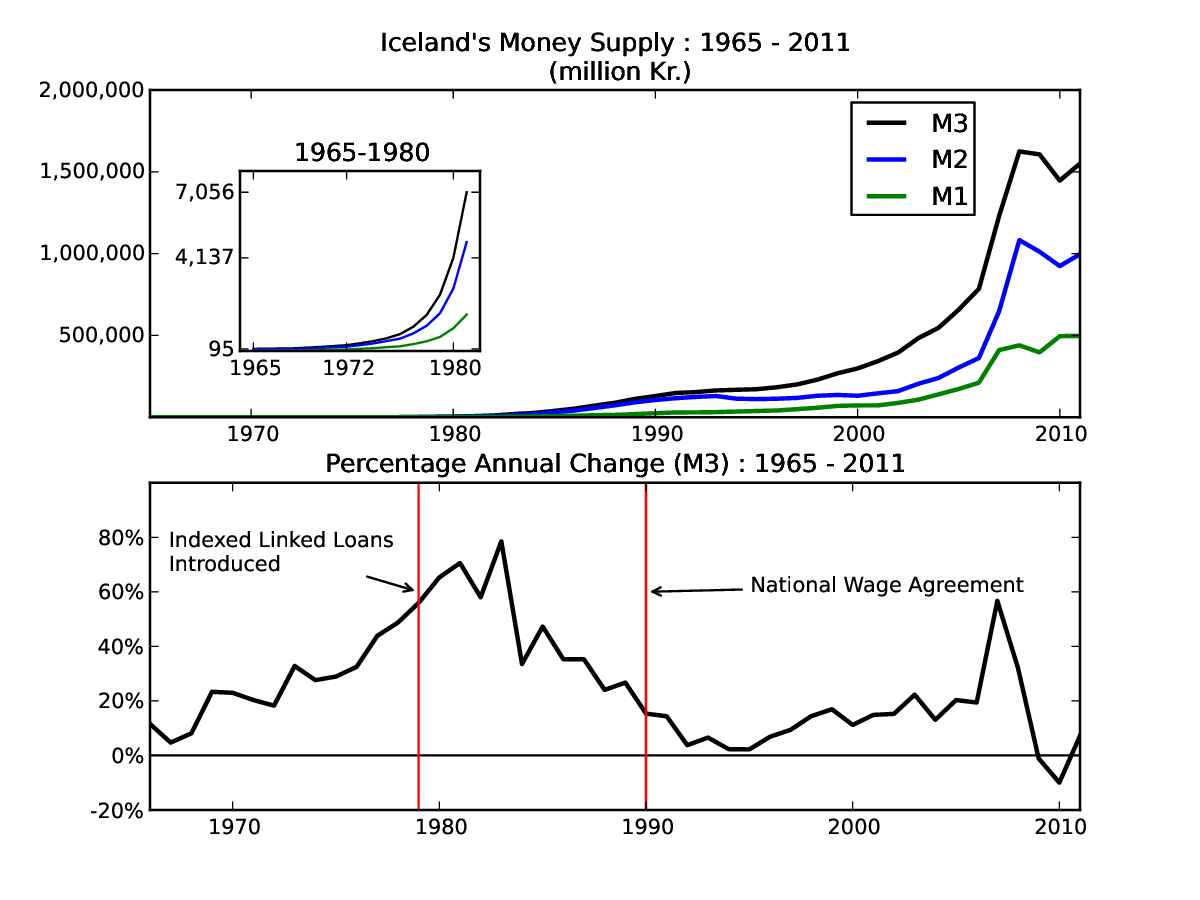}
\caption[]{Iceland M3 Money Supply 1965-2011 and percentage annual 
increase.\protect\footnotemark}
\label{fig:iceland_ms}
\end{center}
\end{figure}

There has also been a long standing question on how these loans interact
with the banking system, and in particular its regulatory controls over
the money and loan supply. Although it is frequently claimed that they
were a key component of efforts to control the hyperinflationary 
environment in Iceland\cite{jonsson.1999}, no specific mechanism has ever been
offered to explain this interaction, and the empirical data
does not support this claim. As shown in 
Figure \ref{fig:iceland_ms}, the peak of hyperinflationary monetary
expansion in Iceland occurred in 1983, 3 years after the loans 
were introduced, and the sharp subsequent falls in the monetary expansion 
rate can be correlated directly with negotiated agreements
to suspend wage indexation. This claim is also contradicted by
the renewed rates of high monetary expansion that were experienced in Iceland
at the end of the 1990's, which ultimately culminated in another 
hyperinflationary
episode in the years immediately preceding the 2007 collapse, despite
the then dominance of these loans within the monetary system.
The second hyperinflationary period in 2005-7 also occurred despite
the Central Bank's adherence to economic theories on banking system
regulation and progressively raising 
interest rates to 18\% to control the 'overheating' of the economy.
\par
\footnotetext{The 1965-1980 
hyperinflationary period is shown 
separately in the inset due to the graphing issues with the
exponential rate of increase.}
\par
The negative amortisation applied to the principal by the indexation on
these loans results in its substantial growth over the period of the loan, it
also raises the question of whether these loans can influence
the monetary expansion rate of the banking system, and through that the
inflation rate?
The mechanical operation of the Banking System is performed using
double entry bookkeeping, and its rules require that a 
matching entry in the bank's 
accounts must be made to balance this principal increase. When a
bank loan is issued the matching entry is the deposit created
on the bank's ledger and added to the borrower's account. While this 
creates money in the form of a bank deposit, the banking system
is in theory regulated so that the creation of bank deposit money
by lending is matched by its removal when loan principal is repaid.
As indexation increases the principal of these loans, two critical
questions must be posed: what is the operation that accompanies
the increase in the loan's principal to maintain the balance of the 
bank's bookkeeping ledgers, and can this operation influence
the money and credit supply originating from the bank? 
\par
Most macro-economic analysis of credit crises concentrates on 
high level statistics of the economy such as overall borrowing levels,
currency relationships and GDP based comparisons, and has a marked
tendency to treat the banking system and its effects on inflation through 
the money and loan supply that it controls as a black box. 
Even recent analysis of the behaviour of these loans is notable for a distinct 
tendency to treat inflation as a force of nature,
and concentrates on indexation as an approach to deal with its consequences, 
rather than discussing the causes of inflation and in
particular the underlying growth of the money supply.  Both 
Mixa\cite{mixa.2010} 
and Finnbogad\'ottir\cite{finnbogadottir.2010} for example, stress
indexation's importance as a means of compensating for Iceland's 
historically high inflation rates, without any exploration of its causes.
\par
In this paper we
will take the opposite approach, and concentrate on the causes of the 
historically high rate of growth in the Icelandic money supply, and the 
behaviour of the banking system that led to this growth.
We will focus on the mechanical causes of the 
regulatory issues encountered in the Icelandic banking system over the
last 30 years, and the general failure of modern banking regulatory
frameworks in Iceland and elsewhere to control the monetary expansion of 
the banking system. We will show that rather
than preventing inflation, the index-linked loans contribute to the
underlying rate of monetary expansion at an amount that is a function of
the rate of inflation, thereby creating a positive feedback
loop operating directly on the money supply.\footnote{Positive feedback 
is a mathematical process within dynamic
systems where the result of a repeated sequence of operations feeds back
into the inputs to the process, resulting
in an increasingly larger output from the process over time, which 
is then re-introduced into the calculation as one of the inputs.}
\par
\section{Ver{\dh}trygg{\dh} l\'{a}n - Index-linked Loans}
\subsection{History}
The Icelandic Krona was first established as an independent currency
by the Money act of 1875, part of a series of constitutional
reforms in 1874 which began the process of establishing independence 
from Denmark. As a de facto member of the Scandinavian 
Monetary Union\cite{bergman.1993}, the exchange rate for an 
Icelandic Krona was set at 
the common exchange rate of 0.4032258g of fine gold to one 
Krona\footnote{The fineness of a previous metal refers to the
ratio of the primary metal to any additives or impurities 
traditionally expressed as parts per 1,000. A fine ounce of
gold is a troy ounce of pure gold content in a gold bar.}
\par
The Gold Standard, which in one form or other
was the dominant banking regulation from the 19th century until the 
collapse of the Bretton Woods agreement in 1973, attempted to regulate the bank
deposit expansion process by creating a fixed relationship between
the price of gold and initially the amount of paper currency issued by
individual banks.  As over time the central bank role became established, 
the regulatory relationship shifted to deposits owned by the commercial
banks which were required to be held at the central bank, and direct control 
over the physical currency was removed from the commercial banks.
\par
Iceland declared independence 
from Denmark in 1944, following which it was a signatory member of the Bretton
Woods fixed exchange system\cite{bretton.1944} until its breakup in 
1973.\footnote{ Data on the Icelandic Money supply since 1965 is available 
from Se{\dh}labanki \'Island's annual report, which is available online 
from the Central Bank's site \url{www.cb.is} from 
1997 and from the Icelandic National Archives from 1965.
The Central Bank provides three measures,
M1, M2 and M3, where M1 is defined as demand deposits and Notes and Coins 
in circulation, M2 as M1 plus sight deposits, and M3 as M2 plus time 
deposits. In this paper we use M3 as the measure for the total Icelandic money
supply. Care has to be taken in international comparisons
of M measures, since there is as yet no commonly agreed definition, and
many countries include forms of debt in their gross monetary measurements.}
There is some evidence that a tradition of centrally controlled
interest rates, rather than those determined by a free market 
process, may have been inherited from its period as a colony. Under 
Danish rule a maximum interest 
rate of 4\% on mortgages was set by law in 1855, this 
limit being 
increased to 6\% in 1890, and 8\% on all interest rates from 1933. Formal 
separation of a central bank did not occur until 1965, prior to that 
Landsbanki provided centralised facilities and in 1952 its interest 
rate was made the national norm\cite{jonsson.1999}. 
\par
Iceland was a signatory member of the Bretton Woods agreement in 1944,
and participated in that regime until its break up in 1973. While the
earlier monetary history appears to be unavailable, high rates of 
monetary expansion can be seen from the beginning of the central 
bank's money supply series in 1965 as shown in Figure \ref{fig:iceland_ms}.
By 1969, and while still in the Bretton Wood's fixed exchange rate framework
the M3 money supply
was growing at an annual rate above 20\%.  By 1973, the year of the
Bretton Woods collapse, the annual rate
of monetary expansion had reached 33\%. Iceland then experienced 
hyperinflationary rates of monetary expansion throughout the 1970's, 
peaking in 1983 when the money supply expanded by 83\% in a single year. 
Understandably this long period of monetary instability has had significant 
repercussions for both the Icelandic economy, and the behaviour of its 
consumers and businesses.
\par
During the 1960's and 1970's Iceland had a central bank reserve requirement
on deposit liabilities of 20\%. At the beginning of the hyperinflationary
period there does appear to have been an attempt to control the expansion 
by increasing the reserve requirement, which was raised
to 28\% with an additional 2\% special liquidity requirement. 
As described in the Icelandic Central Bank's
1981 Annual report\footnote{Page 10.}, this attempt failed:
\begin{quote}
"The monetary development in 1980 supported a theory saying
that fluctuations in the monetary creation are entirely reflected
in the bank's liquidity since no absorbing instruments exist.
The intended contractionary impact of reserve requirements has been
wiped out by the rules of rediscounting"
\end{quote}
The assumptions behind the theoretical description of the 
operation of reserve requirements presented in economic textbooks ignore the 
practicalities involved in the 
day to day management of temporary imbalances that are created
as money flows between banks. Inter-bank lending and other
methods such as rediscounting\footnote{Borrowing based on an
underlying asset which is itself a debt.}  allow banks to 
manage their day to day reserve requirement, by lending excess
reserves and borrowing to make up shortfalls, and this also
allows them to circumvent the full force of the reserve controls.
This problem was also mentioned by
Keynes in 1929\cite{keynes.1929}\footnote{Pages 234-243.} in a 
discussion on the 
effects of the different practices between the US Federal Reserve Banks, and 
the Bank of England in managing the bank rate and rediscounting.
\par
The immediate cause of the 1969-1988 hyperinflation\footnote{Defining
hyperinflation somewhat arbitrarily as an annual monetary expansion greater than
20\%.} appears to have been physical printing
of money by the Icelandic Government as a source of revenue. Although
direct proof of this is hard to obtain, an International Monetary Fund(IMF) 
review of the decade following the introduction of indexation by
Cornelius\cite{cornelius.1990} gives it as the cause, and it is consistent with 
the subsequent behaviour of the monetary system. A 1998 working
paper from the Icelandic central bank by Andersen and 
Gu{\dh}mundsson\cite{andersen.1998} alludes to
seignorage\footnote{Strictly, the difference between the value of 
a monetary token, and the cost of its production. Used in Economics
as a general term for the government's ability to use its privileged
position to profit from printing money.} as a cause but then 
suggests that this was a relatively 
small contribution as a percentage of GDP (3\%). This is a slightly
disingenuous argument as it overlooks the role of physical money in
the regulatory control over the behaviour of the banking system.
The amount of lending and consequent bank deposit 
creation by commercial banks in 
gold standard regulatory regimes\footnote{Under the current Basel 
regulatory framework, there is
still a multiplier relationship, but it is significantly throttled by 
the separate regulatory controls on each individual
bank's capital reserve requirements.}
is partially regulated as a multiplier of their reserves at the central 
bank which can be increased by cash deposits. Increases
in reserve holdings caused by physical printing can trigger a considerably 
greater expansion in the part of the money supply represented by deposits
within the banking system, in the absence of any other regulatory controls
on the banking system.\footnote{The relationship between physical
cash, a bank asset, and the customer's deposit at the bank, is 
statistically multiplexed within the banking system.}
As a consequence, attempts by Governments within fractional reserve 
banking regimes to profit from money printing
necessarily fail, as the purchasing power of physical money introduced into the 
monetary system by the government is rapidly reduced by the considerably
larger expansion it creates in 
bank deposits. The resulting hyperinflation has however 
proved to be one of the most economically damaging forces in human
history.
\par
The Icelandic monetary authority's response to the 1980's hyperinflation 
as described by J\'{o}nsson\cite{jonsson.1999} in Central Bank Working 
Paper No. 5, was widespread indexation, linking prices and wages directly 
to the inflation rate. A side-effect of the hyperinflation was
a short window of opportunity for those with long term loans taken out
before the hyperinflation began, whose loans were effectively written
off by the extreme monetary expansion at the end of the 1970's. 
J\'{o}nsson's  paper minimises attention to the precise causes of the 
hyperinflation, and concentrates on the indexation of lending 
as a solution to the particular problem of preserving the value of loan 
capital.  This solution was the introduction of indexation on loans
loans in 1979, with the value of loan capital being directly 
linked to the CPI inflation rate.
\par
In an environment where the money supply is doubling every two years 
and the consumer inflation rate is over 50\% a year; where wages and 
prices are largely
indexed to the inflation rate; it is entirely understandable that
protection of loan capital should be seen as a priority. Moreover, as 
J\'onsson points out, indexation of debt provided a critical pressure
on political efforts to stop the hyperinflationary expansion as increases
in wages due to indexation were then matched by increases in borrowing costs,
so a behavioural component can certainly be argued for.
Unfortunately the introduction of indexation on bank lending disregards 
several critical 
differences between bank loans, and other forms of lending such as 
corporate and government bonds. In particular, it neglects to consider
the role the banking system itself plays in creating inflation, through
the expansion of the money supply in the form of bank deposits. It also
overlooks the source of bank profit, which is derived from the interest
on the loans they make. Because bank lending relies on the creation
and destruction of money (as principal is repaid), while profits are
derived from the interest payments on outstanding loans, banks are relatively
immune to the effects of any money supply expansion that they cause
and do not experience capital destruction in the way that
other lenders do. 
\par
The hyperinflationary period continued throughout the 1980's as various
attempts were made to unwind the set of mutual feedback relationships
that had now been created between the money supply, lending, and indexation
on wages. The monetary expansion rate dropped noticeably in 1984 following the
introduction in 1983 of a temporary suspension of wage
indexation\cite{andersen.1998}, and was finally brought under control
in 1990 when the national wage agreement ended wage indexation with
monetary growth falling sharply from 14.36\% in 1991 to 3.77\% in 1992.
\par
Loan indexation remained however, on mortgages, some forms of commercial
borrowing and student loans; and some of its longer term
consequences were becoming apparent. In its 1992 Annual Report the Central
Bank reports that:
\begin{quote}
"Third, the indexation of financial assets as well as higher and
positive interest rates have had the impact that household debt
has accumulated instead of being eroded through inflation."
\end{quote}
Other consequences went unrecognised. The 1990's were
a period of low inflation, in Iceland and elsewhere, as the impact of
rapid technological developments in manufacturing
substantially increased the supply of goods and services worldwide. 
Iceland experienced historically low rates of monetary expansion 
between 1994 to 1998 which coincided
with changes to the banking regulatory framework. The new Bank of
International Settlement (BIS) rules on the treatment of equity capital,
generally known as the Basel Accords\cite{basel.1988}, came into effect in 
Iceland at the beginning of 1993. They introduced a number of significant
changes to the regulatory framework as described in the Central Bank's
1992 report:
\begin{quote}
The BIS rules on equity capital came into effect at the beginning
of 1993. They are somewhat different from the rules that have
applied in this country up to now, the main difference being
twofold. One, the amount of equity needed to cover loans differs,
depending on the type of collateral behind a loan. Thus, no
equity need stand behind a loan to the central government for
example, compared with 8\% equity against loans to companies without
a quality mortgage. Two the definition of equity ratio is
widened to some extent so that certain subordinated debt of banks
may be counted as part of equity. Furthermore the equity ratio
is increased from 5\% to 8\% of capital which need not mean a
substantial increase since the percentage is calculated on a
lower base than before.
\end{quote}
The Basel Accords formalised a secondary control on
the regulatory limits on bank lending, the capital reserve requirement,
and were primarily intended to control the exposure of banks to excessive
loan defaults by regulating the amount of risk they could be exposed to.
They were not intended to directly regulate the bank deposit
expansion process, and that they do in fact exert an influence
on it, is probably serendipitous.
While the central bank reserves regulate
the amount of lending that a bank can perform as a function
of those deposits for which it is required to maintain a
reserve\footnote{This varies considerably, for example in the USA
only "Net transaction accounts" incur a reserve requirement. In
Iceland currently reserve requirements apply to all accounts with a maturity
period of less than 2 years, as well as debt securities and money
market instruments. Rules on Minimum Reserve Requirements, 15/4/2008,
\url{http://www.cb.is/lisalib/getfile.aspx?itemid=5850}}, the capital
reserve requirement regulates the amount of lending that a bank can perform as
a risk weighted function of its capital base.
\par
Over time the capital reserve requirement generally dominates
in determining a bank's lending limits. In the event of a shortfall
in their central bank reserves, Banks can borrow either
from other banks or as a last resort from the central bank. When problems
arise due to pressure on clearing and reserve liquidity the central
bank can generally be relied on to intervene.
As a consequence, the central bank reserve requirement is ineffective
in limiting long term lending expansion, and this can be seen quite
clearly in the Icelandic statistics where the monetary expansion during the 
2000's was accompanied by
a sharp increase in the amount of inter-bank lending.
\par
Banks can individually increase the size of their capital reserves 
from profits, and so the capital requirements place no absolute
limits on money and loan expansion.
By any standards the index-linked loans are also extremely profitable
for their issuers. They carry a guaranteed base rate which is
calculated on a negatively amortised principal controlled by 
indexation to the CPI. In addition personal debt in Iceland has historically 
been treated as full recourse, and in practice
cannot be discharged through bankruptcy. However the high profitability
of the loans depends entirely on a positive inflation rate. As shown in 
Figure \ref{fig:vl_projected_40_0}, for rates of inflation that are 
approximately
3\% or less (it depends to some extent on loan duration), the loans behave
similarly to fixed rate compound interest mortgages, and the negative
amortisation isn't triggered. Had the
banking system's regulatory framework in Iceland succeeded in imposing
an absolute limit on deposit growth, these loans could not have caused
an increase in the money supply and the consequent inflation. In that regime, 
any increase in the bank deposit portion of the money supply
caused by the negative amortisation would have had to be offset
by a decrease in lending. In practice, the net effect would have
been stable house prices, a stable or decreasing CPI, and the loans 
would have been no more expensive than their fixed rate equivalents.
\par
With no absolute limit on the amount banks can increase
their capital holdings, except for their profitability, this did
not occur. Instead, as banks recognised the additional income
accrued from the negative amortisation of the index-linked
loans, they were able to use part of this income to increase
their required capital holdings. This in turn allowed them
to increase the size of their loan book, resulting in an increase
in the total amount of money represented by bank deposits, and as
a consequence, in additional inflation. Financial liberalisation
in the late 1990's only exacerbated the problem by fuelling a credit
boom in housing, which since the Icelandic CPI calculation includes
a house price component, further accelerated the monetary growth from
these loans.
\subsection{Detailed Description}
Indexed linked loans are structured with two components:
a base interest rate, and the indexation which is applied on the principal.
Originally the base rate was fixed for the duration of the loan, but 
since 2004 loans have been available with base rates that are fixed for
5 years and then reviewed. There do not appear to be any contractual 
limits on how this review will be performed.
\par
\begin{figure}[ht]
\centering
\includegraphics[width=8cm]{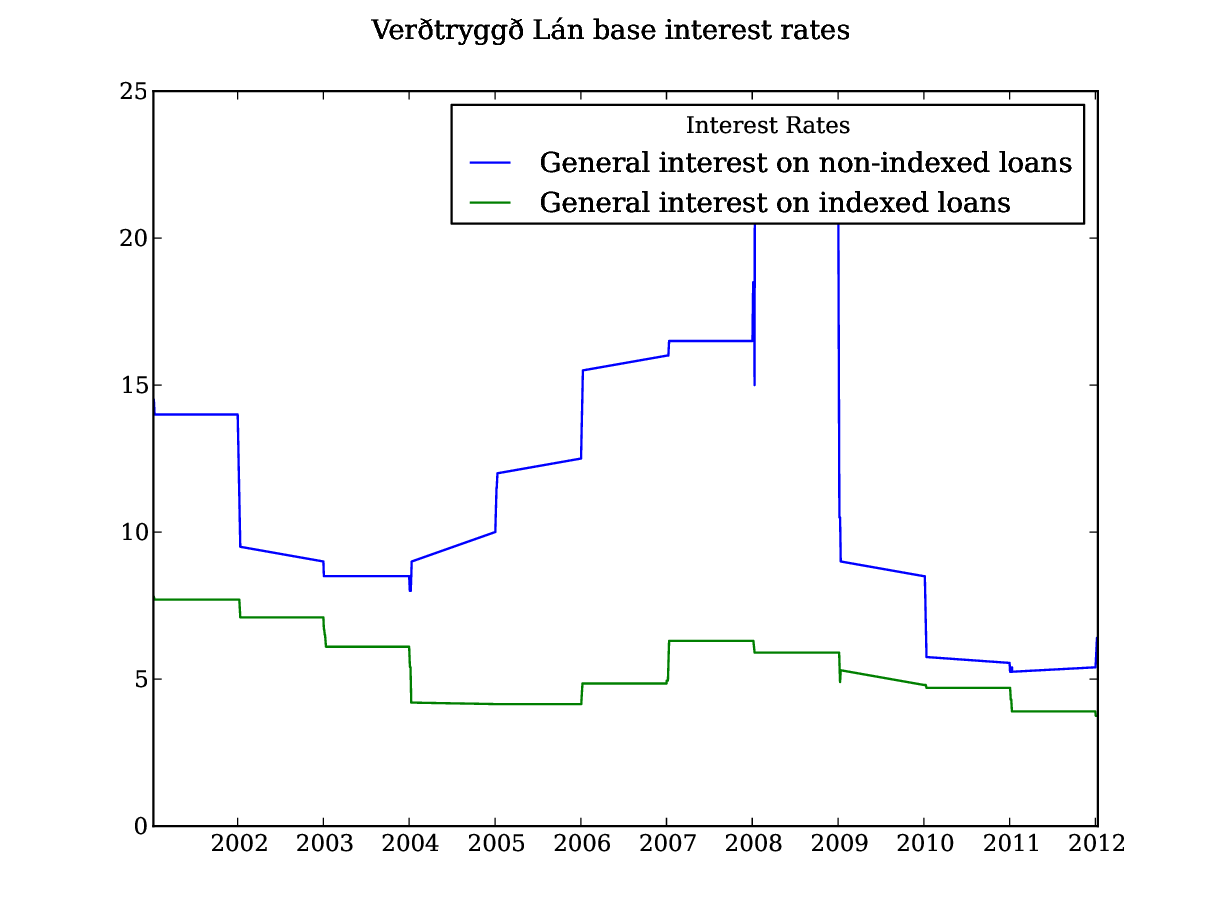}
\caption{Central Bank base rates for lending 2001-2012}
\label{fig:vl_interest_rates}
\end{figure}

Finding a reliable series of the base rate interest rate that has been
used for these loans has been difficult.  Since 2001 a rate has been 
determined by the central bank\footnote{Act No. 38/2001 on interest 
and price indexation, Article 4: "In cases involving indexed claims, 
the interest 
rate shall be equal to the rate decided by the Central Bank, having regard to 
the lowest rate of interest on new, general, indexed loans from credit 
institutions, and posted in accordance with Article 10"}, however figures
from some institutions are at variance with this rate. 
Information on the base interest rate used for these loans in this paper
is derived from the Statistics Iceland weighted average interest
rates series for the commercial banks, "Indexed securities Real interest, 
\% per year available at \url{www.statice.is}".  These have been
cross checked with the Central Bank series from 
2001.\footnote{Source spreadsheet 
Almvex.xls, available at: \url{http://www.cb.is/statistics/interest-rates}} 
which provides the base central bank rate set for these loans. 
They appear
to be slightly lower than the rates used by the commercial banks used, 
prior to privatisation in 2001,
so these graphs may slightly underestimate actual payment levels.
Figure \ref{fig:vl_interest_rates} shows the base rate values for 
indexed-linked loans and non-indexed loans from
the Central Bank's statistics.
\par
\par
The indexation 
component is calculated from the CPI index using the Janus 
rule(see below), which is applied to the outstanding principal of the loan. 
Calculation of the repayments on these loans over their lifetime 
depends on the combination of these two data series.
The most reliable source appears to be Statistics 
Iceland which provides several indexation 
series\footnote{\url{www.statice.is}, Statistics section 8. Prices and 
consumption, Consumer price index, Indices for indexation from 1979.} 
most of which date from the 1990's or later. Table \ref{tbl:index_tbls}
shows the three tables listed under 'Indices for indexation from 1979'.
\begin{table}[ht]
\centering
\begin{tabular}{lc}
\hline
Series  &  \\ 
\hline
Credit Terms Index                                 & 1979-1995 \\
Consumer price index for indexation                & 1995-2013 \\ 
The old credit term index of financial obligations & 1995-2013 \\ 
\hline
\end{tabular}
\caption{}
\label{tbl:index_tbls}
\end{table}
Since the 'old credit term index' series appears to be a continuation of
the 'consumer price index for indexation series' and together they 
cover the entire period, these are the series used in this paper for 
calculations.
\par
Given that the principal of these loans is indexed to inflation, it is 
interesting that the Central Bank also chose to vary the base rate by
significant amounts over the 2001-2012 period, seemingly in response to
increases in inflation during that period and presumably guided
by Keynesian theories. Perhaps more
significant, at least for the theories, is the absence of any 
corresponding contraction in the expansion rate of the money supply during
this period.
\par
\begin{figure}[ht]
\centering
\begin{tabular}{cc}
\subfloat[1981 505,000 ISK @ 2.5\% base rate]
   {\includegraphics[scale=0.27 ]{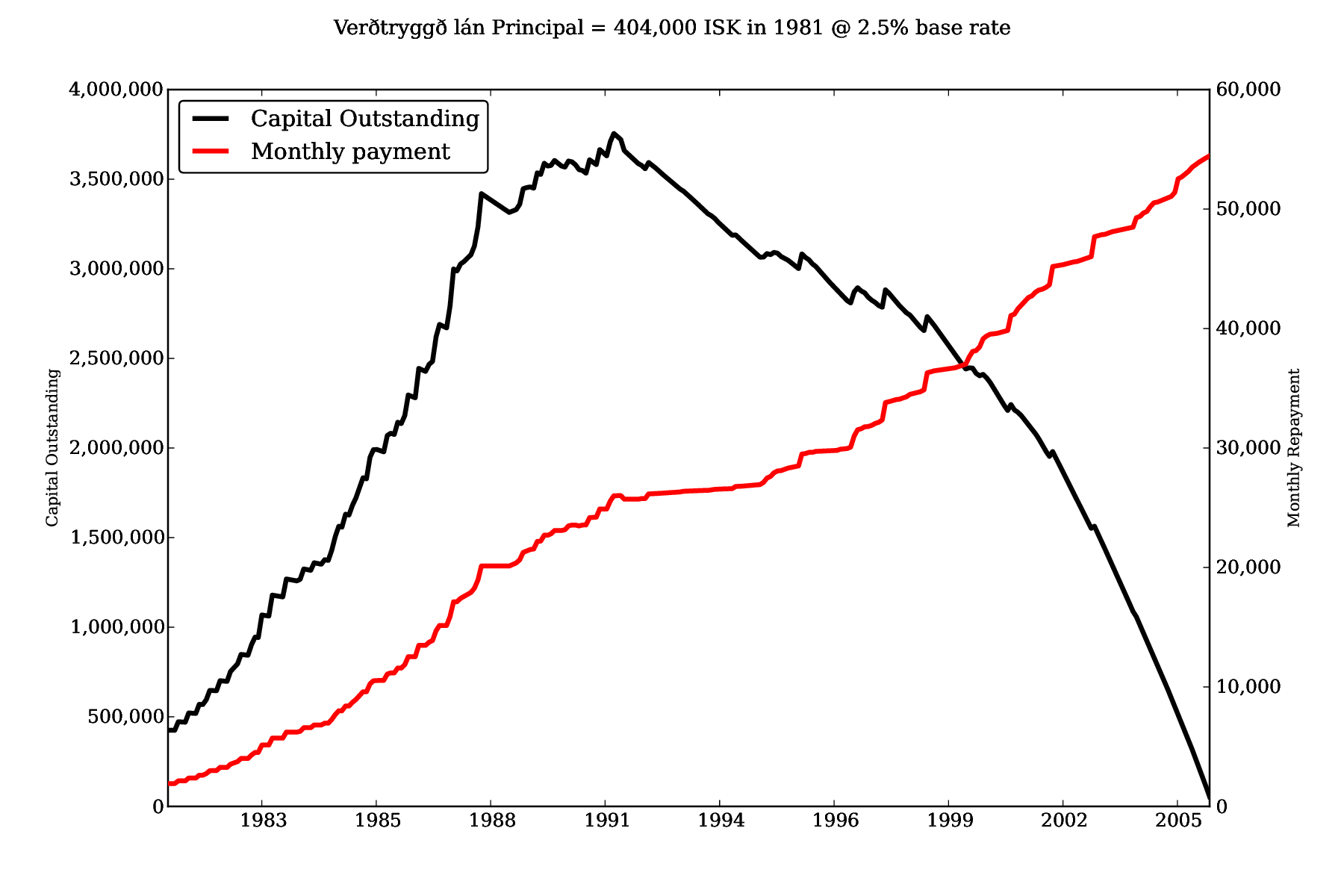}}   &
\subfloat[1985 2,097,000 ISK @ 5\% base rate]
   {\includegraphics[scale=0.27 ]{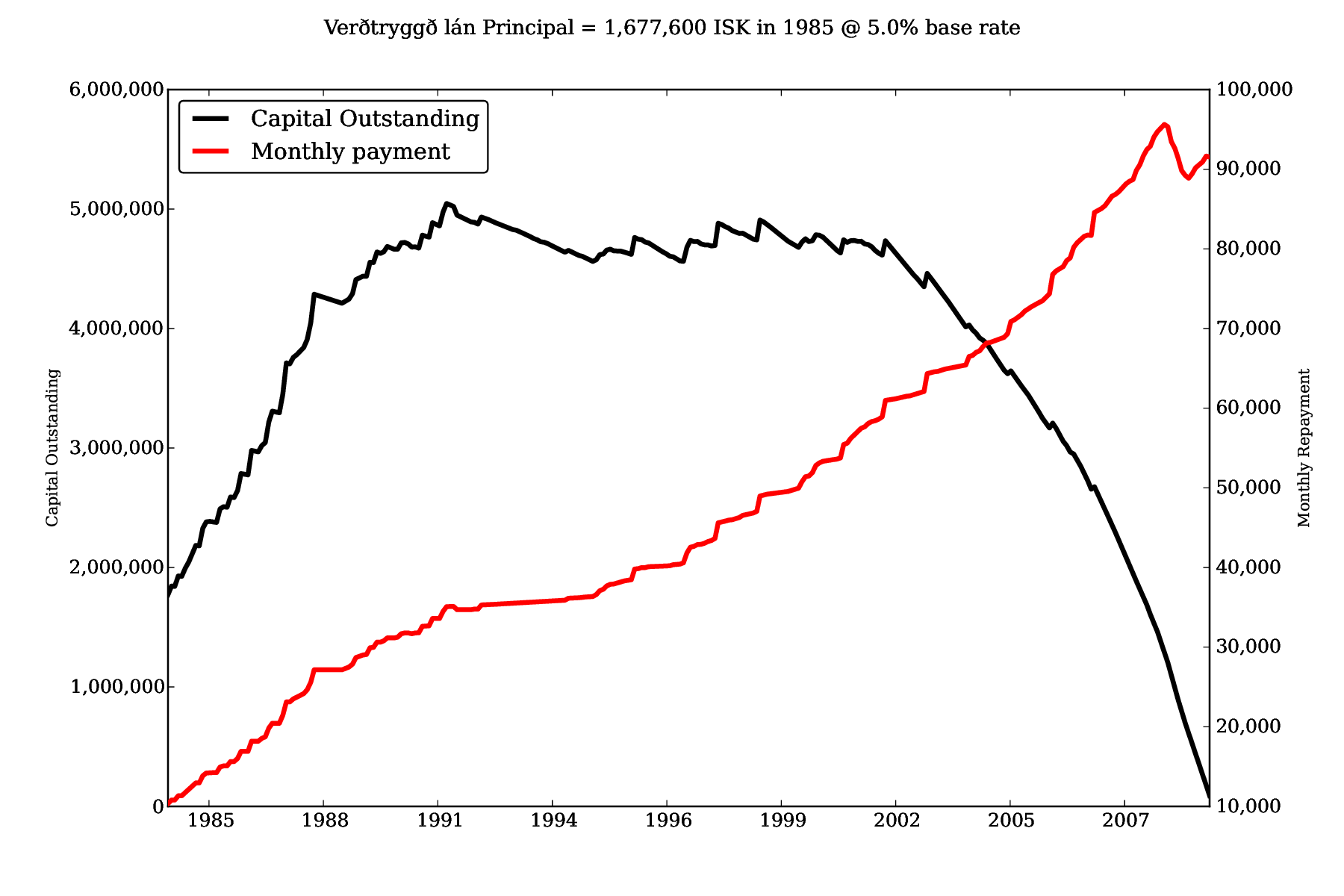}} \\
\subfloat[1990 6,499,000 ISK @ 8\% base rate]
   {\includegraphics[scale=0.27 ]{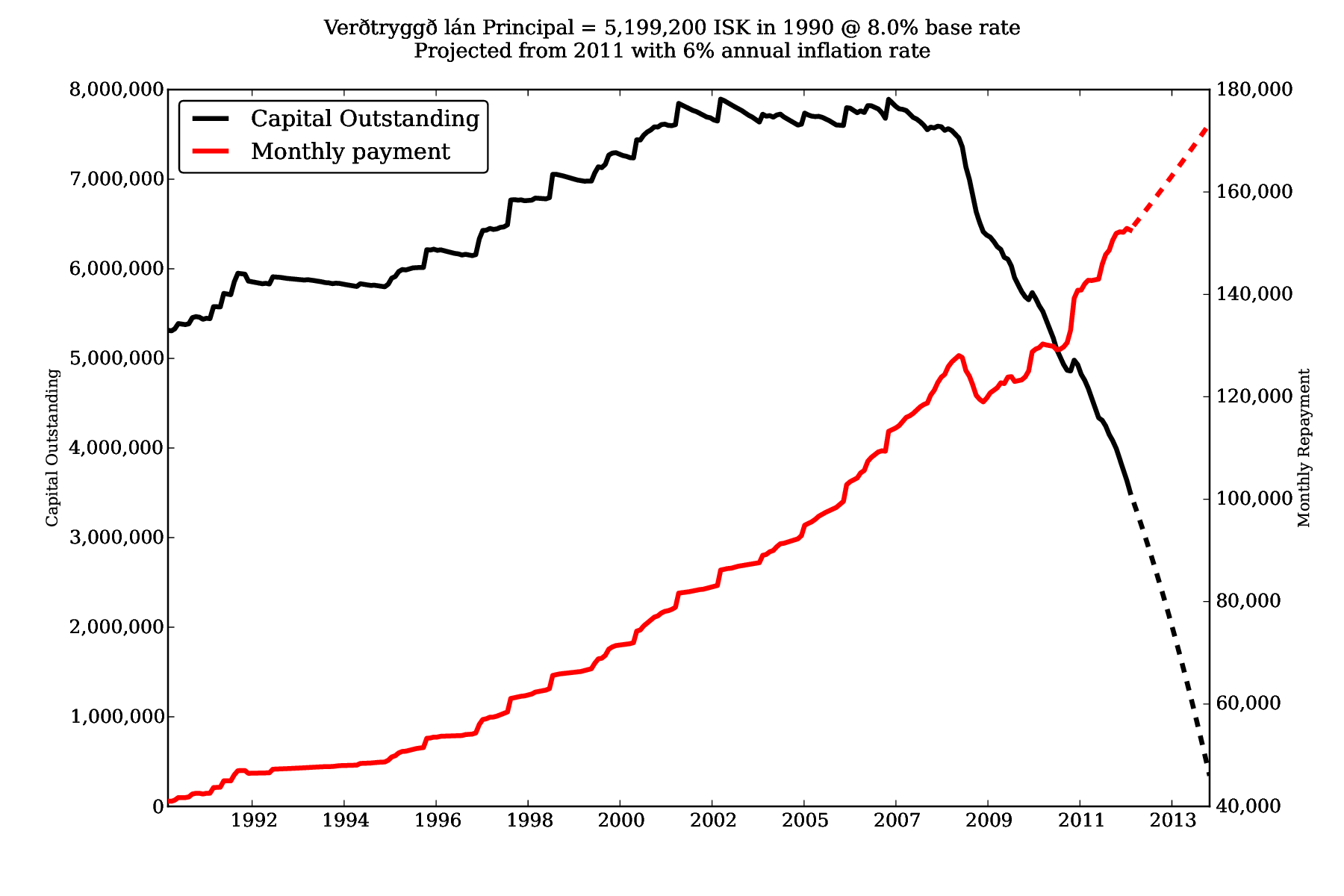}}  &
\subfloat[1995 7,697,000 ISK @ 8.7\% base rate]
   {\includegraphics[scale=0.27 ]{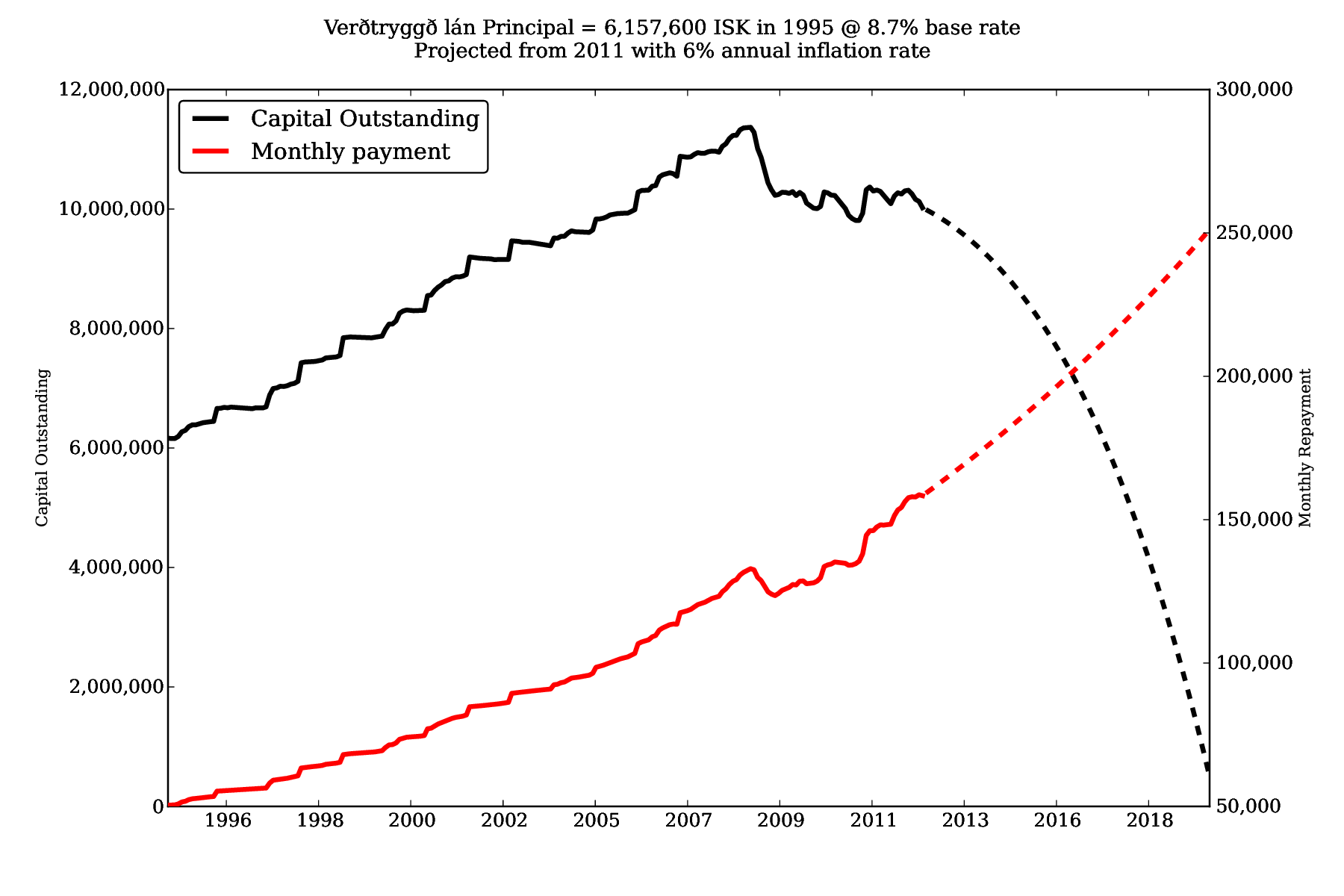}}  \\
\subfloat[2000 11,333,000 ISK @ 7.8\% base rate]
   {\includegraphics[scale=0.27 ]{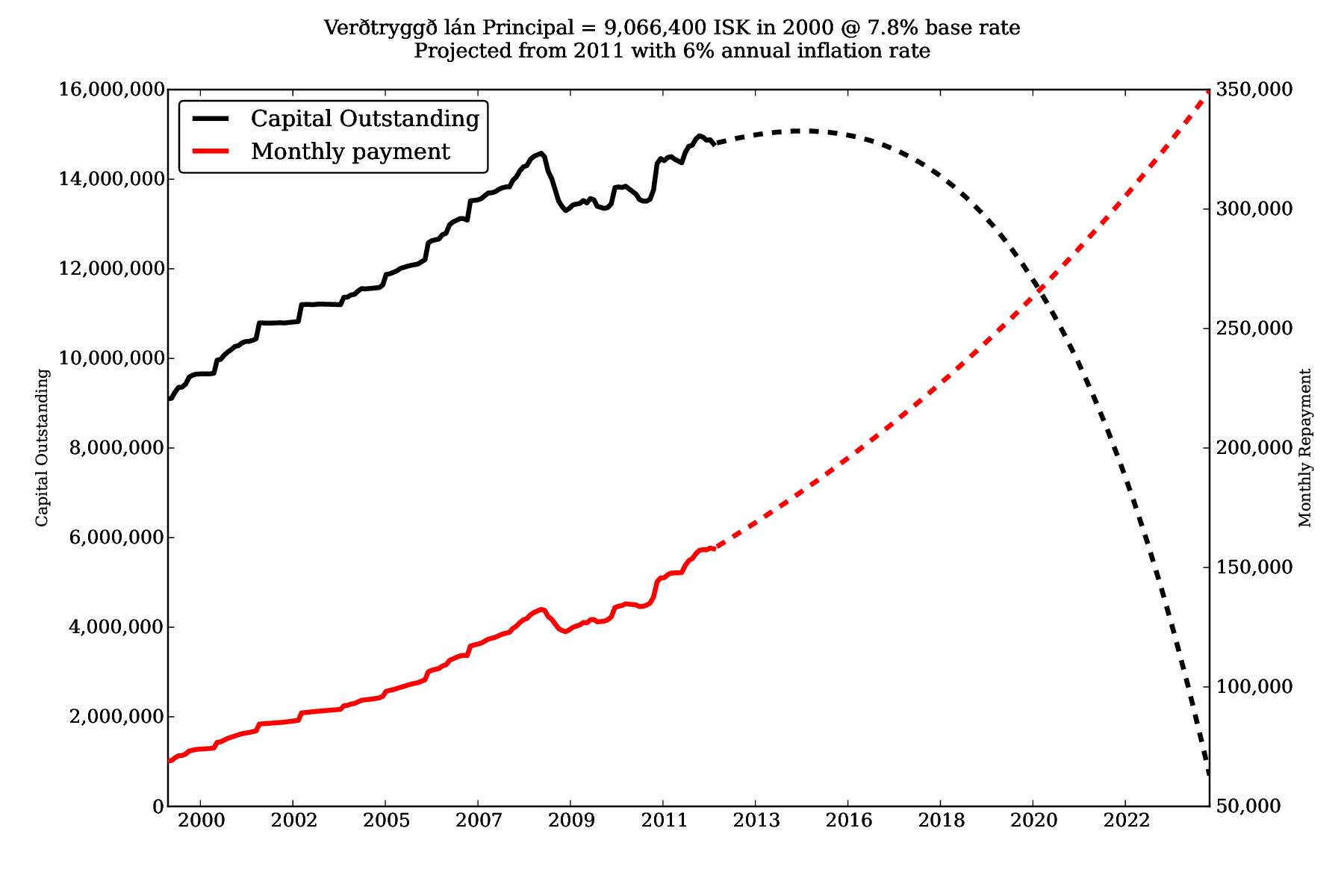}} &
\subfloat[2005 20,779,000 ISK @ 4.2\% base rate]
   {\includegraphics[scale=0.27 ]{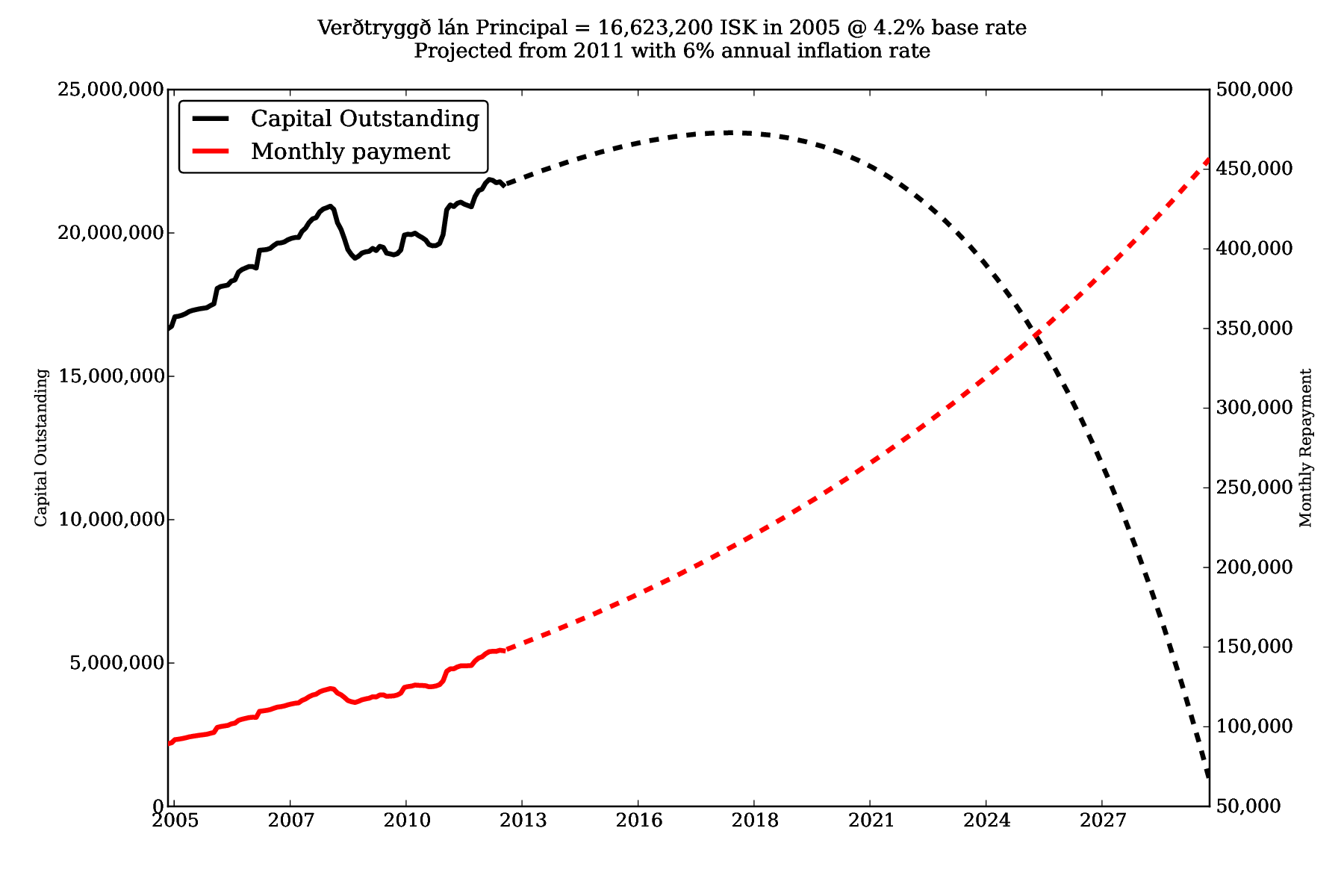}} \\
\end{tabular}
\caption{Indexed Loan profiles for 25 year loans calculated on the basis of 80\% of the average house price in Reykjavik.}
\label{fig:vl_actual_25}
\end{figure}
\begin{figure}[h!t]
\centering
\begin{tabular}{cc}
\subfloat[1981 505,000 ISK @ 2.5\% base rate]
   {\includegraphics[scale=0.27 ]{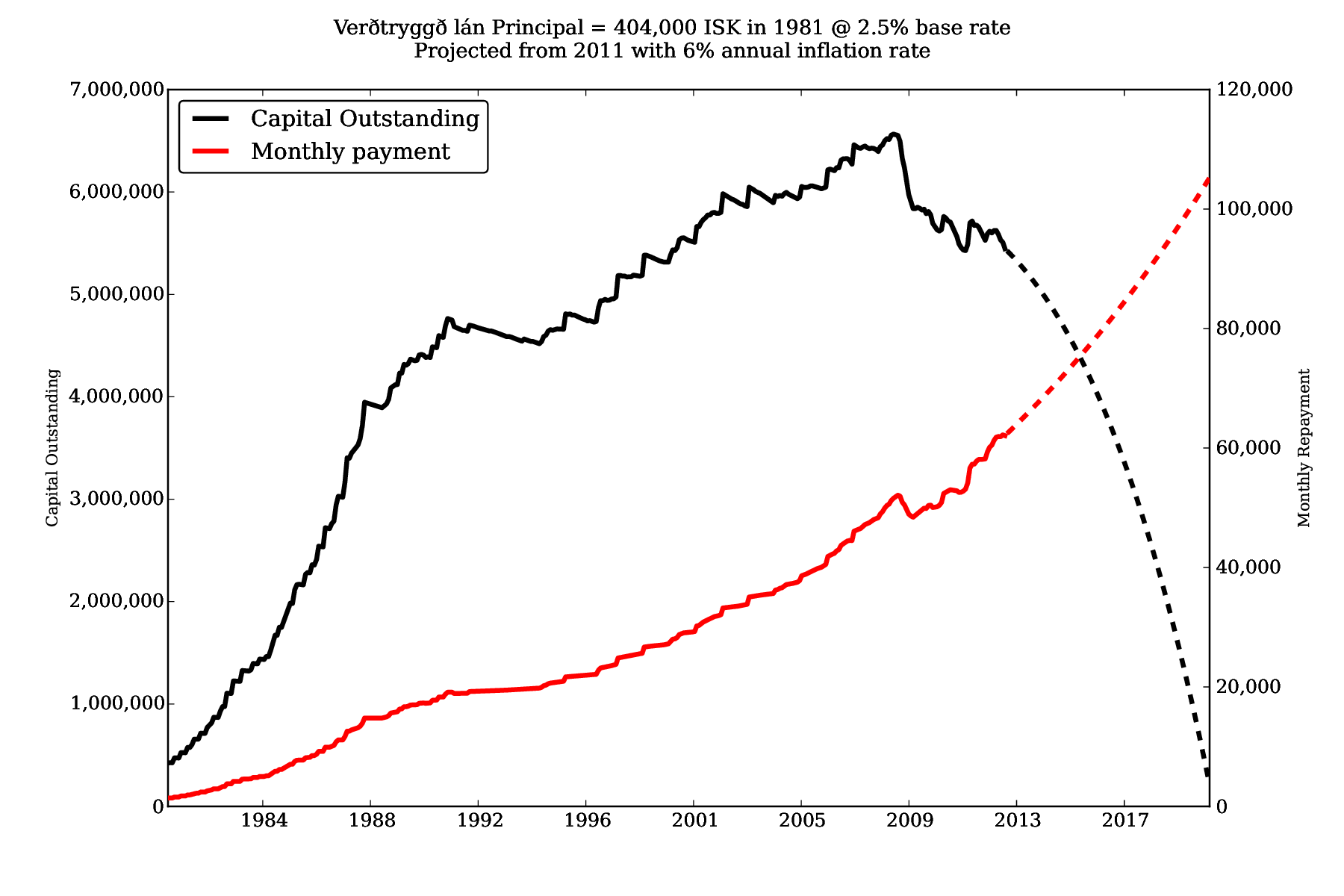}}   &
\subfloat[1985 2,097,000 ISK @ 5\% base rate]
   {\includegraphics[scale=0.27 ]{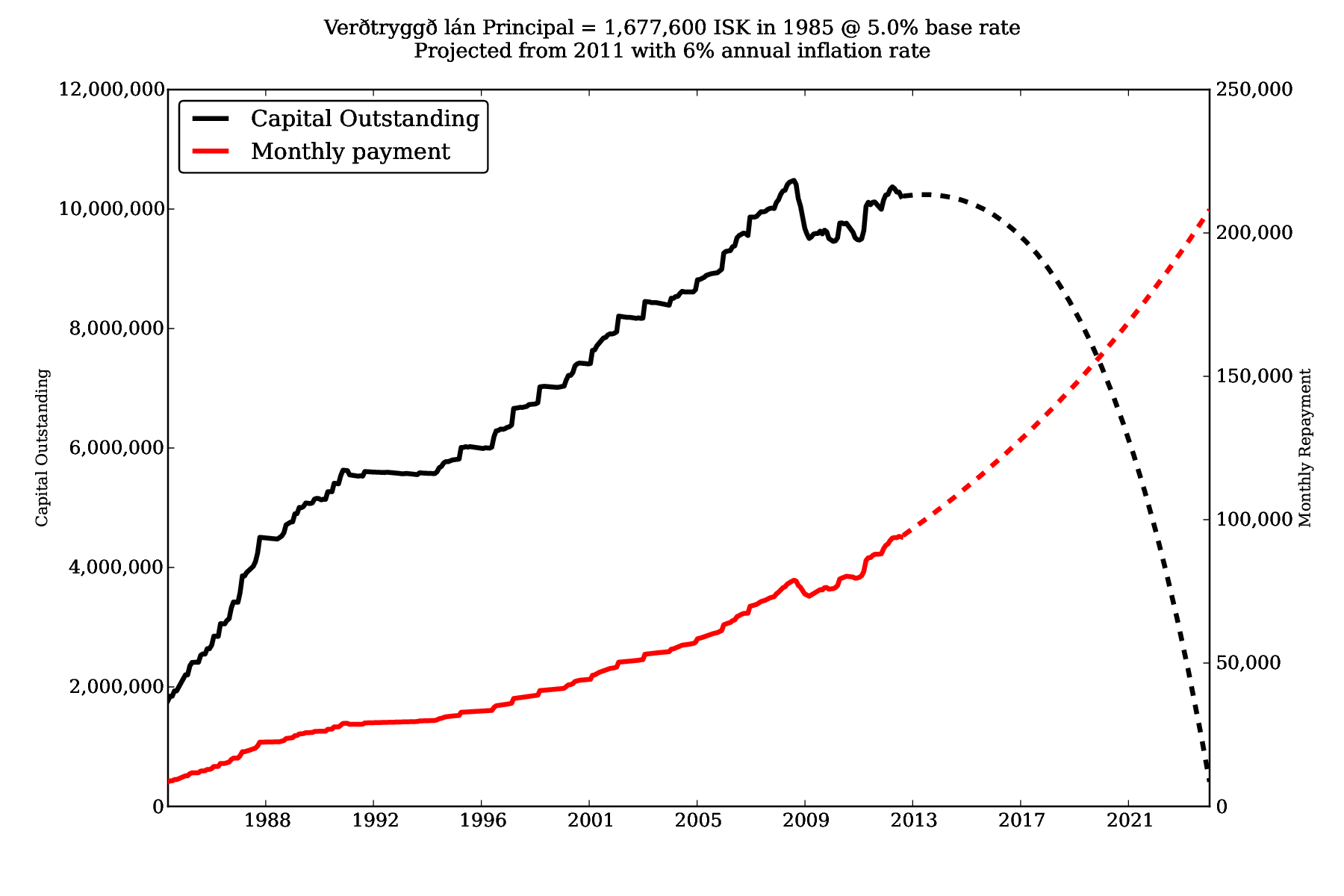}} \\
\subfloat[1990 6,499,000 ISK @ 8\% base rate]
   {\includegraphics[scale=0.27 ]{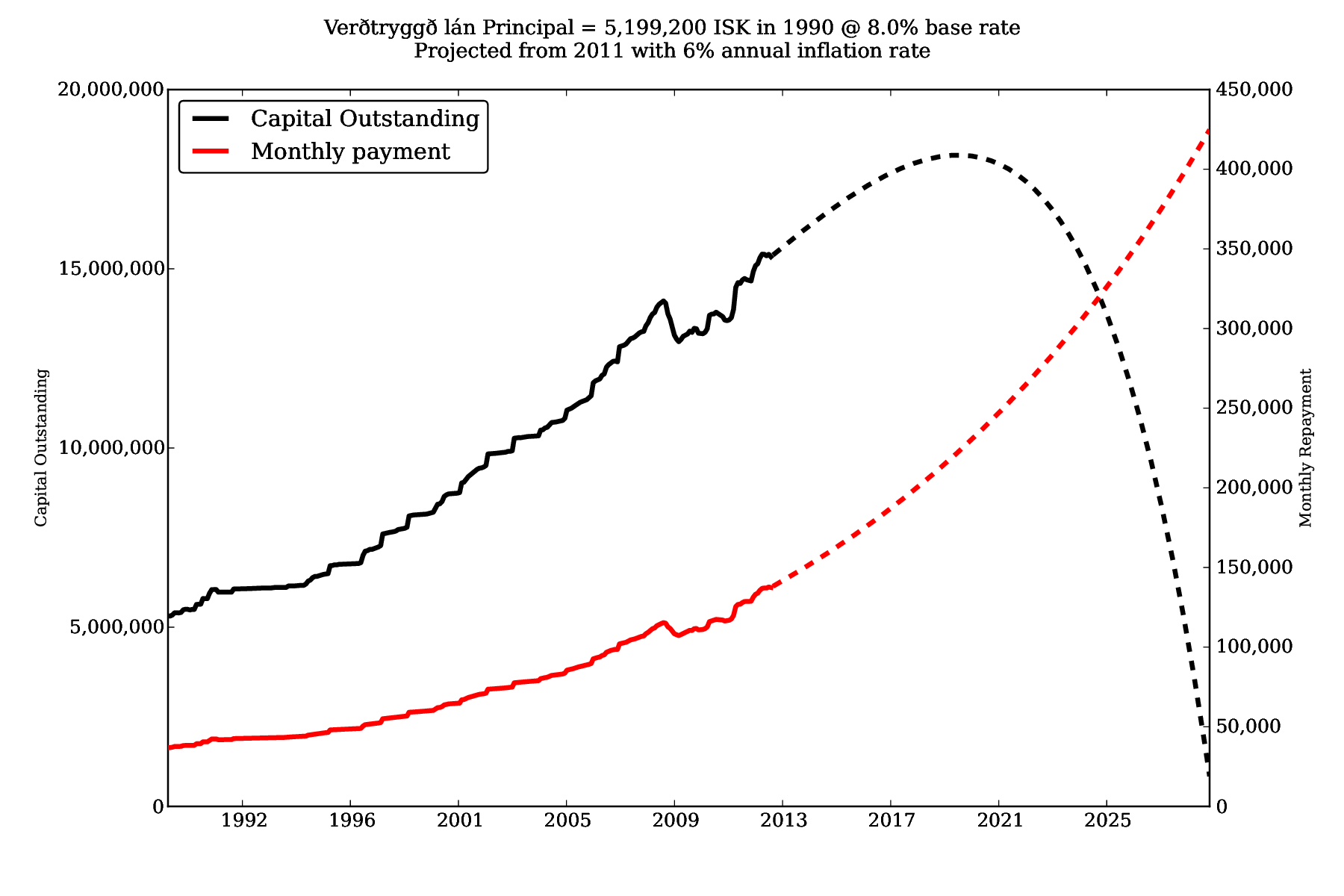}}  &
\subfloat[1995 7,697,000 ISK @ 8.7\% base rate]
   {\includegraphics[scale=0.27 ]{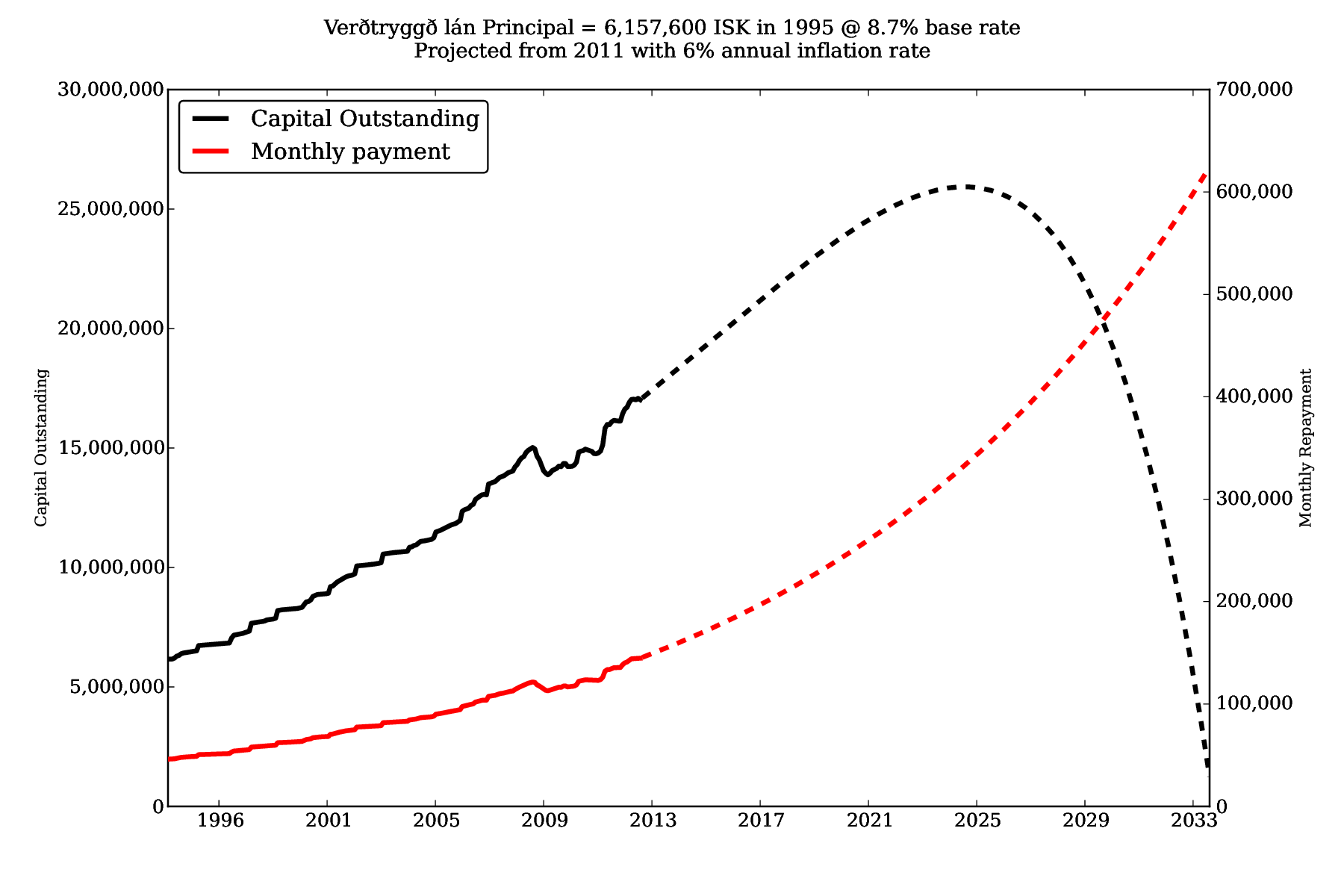}}  \\
\subfloat[2000 11,333,000 ISK @ 7.8\% base rate]
   {\includegraphics[scale=0.27 ]{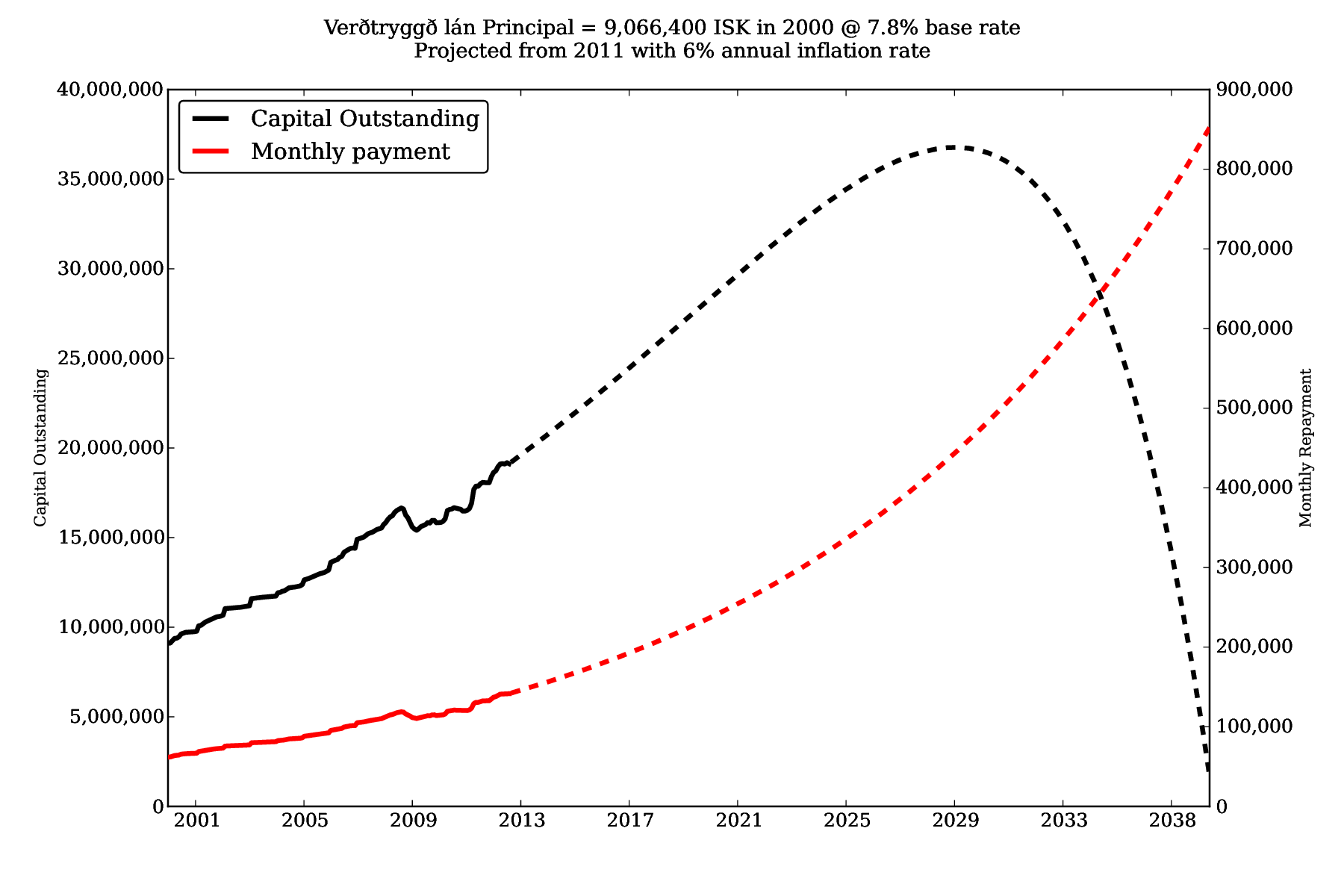}} &
\subfloat[2005 20,779,000 ISK @ 4.2\% base rate]
   {\includegraphics[scale=0.27 ]{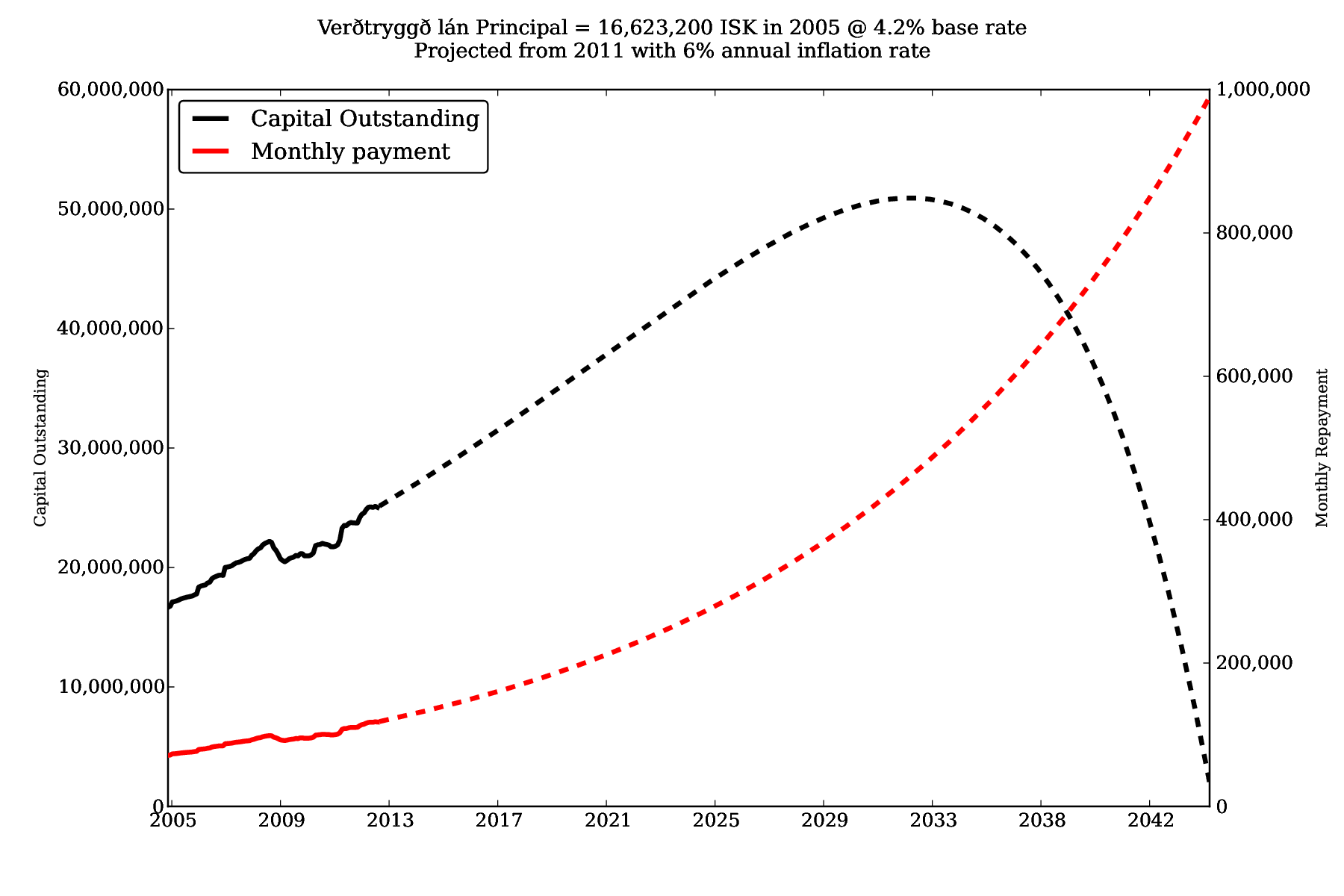}} \\
\end{tabular}
\caption{Indexed Loan profiles for 40 year loans calculated on the basis of 
80\% of the average house price in Reykjavik.}
\label{fig:vl_actual_40}
\end{figure}

The indexation component, which is applied to the outstanding principal,
is weighted across a 12 month period using the Janus 
rule\footnote{J\'{o}nsson\cite{jonsson.1999} p14-15}, which
applies past inflation and a future estimate of inflation to arrive
at a weighted average. The calculation is performed using this approach
in order to mitigate the effect of large short term fluctuations in the CPI, 
but the calculation of successive percentage changes in this way is not
mathematically neutral: as a side
effect it effectively prevents the borrower from benefiting from short
periods of deflation which would have triggered a corresponding decrease
in loan principal.\footnote{Mathematically, the application of successive 
percentage modifications is not commutative, so this treatment results
in the indexed amount of the loan growing slightly faster than it would 
have otherwise.}  Longer periods of deflation can cause principal
reductions, as can be seen in 
Figures \ref{fig:vl_actual_25} and \ref{fig:vl_actual_40}
following the 2008 crash. 
\par
\begin{figure}[ht]
\centering
\begin{tabular}{cc}
\subfloat[25 year loan amortisation]{\includegraphics[width=8cm]{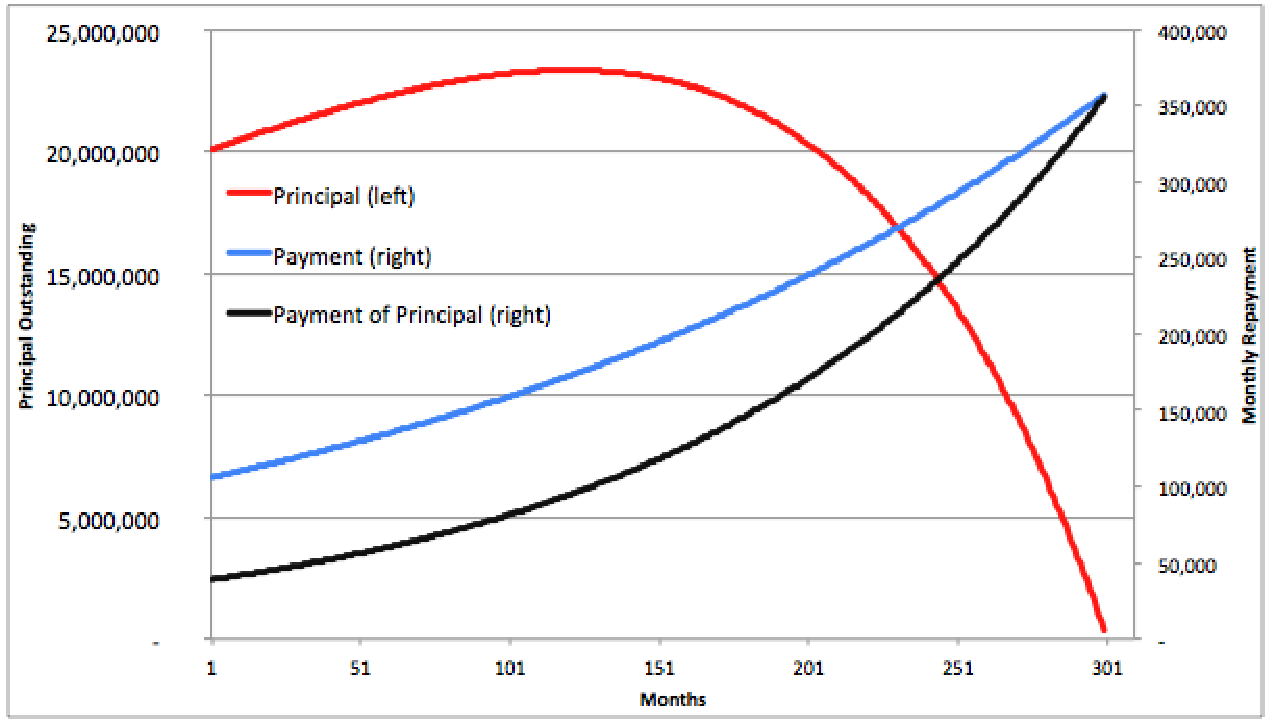}} &
\subfloat[40 year loan amortisation]{\includegraphics[width=8cm]{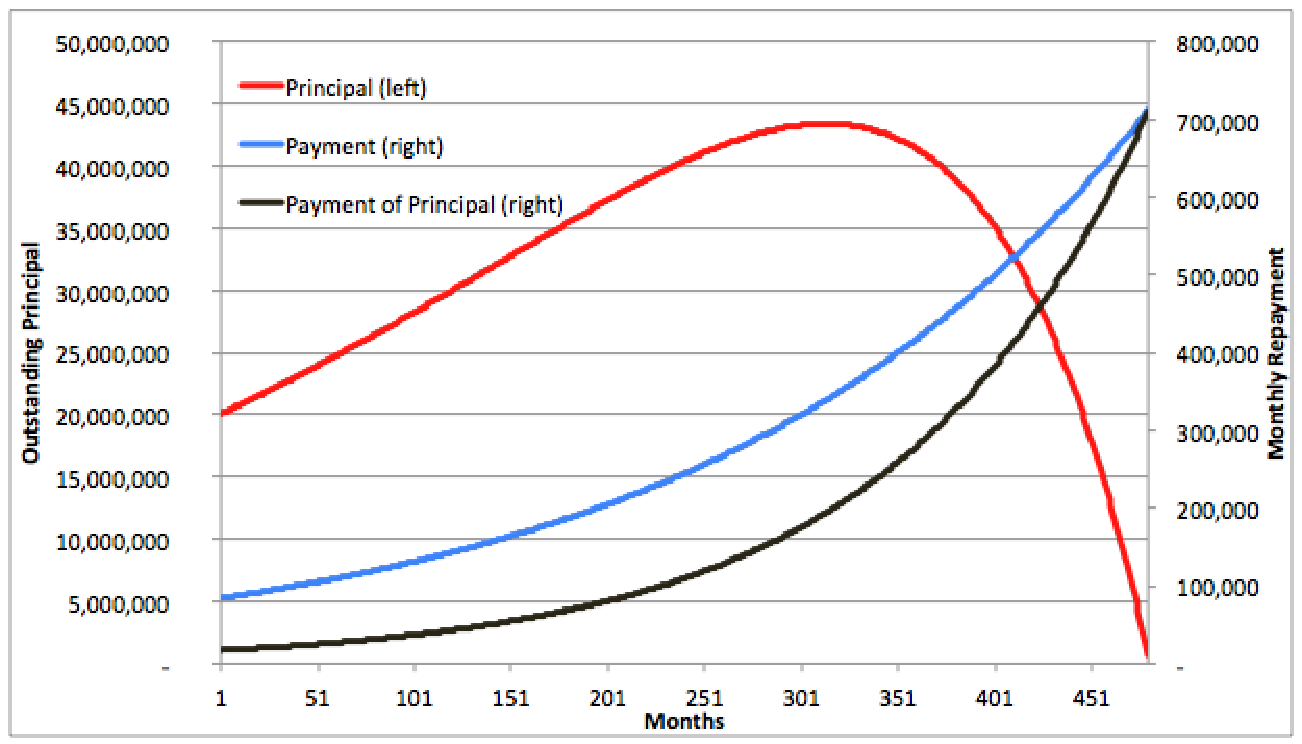}}
\end{tabular}
\caption{25 and 40 year theoretical repayment profiles @ 4\% Base and 5.4\% median CPI}
\label{fig:vl_amortization}
\end{figure}

Figure \ref{fig:vl_amortization} shows the theoretical repayment profile of the 
principal for sample loans with a 4\% base rate, and the median inflation rate
for the 1980-2011 period of 5.4\%\footnote{Since the hyperinflationary period
in the 1980's distorts the average rate considerably, the median 
rather than the average is used for these charts.}.
\par
The repayment structure for the loans results in a
varying period of negative amortisation, the length of which depends 
on the duration of the loan and the rate of inflation during the loan. 
The loans are structured so that the indexation component results in 
principal growth (negative amortisation)
from the beginning of the loan. However borrower
repayments can theoretically overcome the negative amortisation 
during the first part of the loan during periods of very low inflation.
Depending to some extent on the repayment point in the loan,
the CPI must be below approximately 3\% for a 25 year loan and  
1.5\% for a 40 year loan. This behaviour 
can be seen very briefly during the 1990's when between 1994-1998
the average inflation rate was between 1.5-2.25\%. 

Figures \ref{fig:vl_actual_25} and \ref{fig:vl_actual_40} show a 
series of repayment profiles calculated from the indexation figures from 
Statistics Iceland\footnote{Source: \url{www.statice.is} statistical series
"Indices for indexation from 1979"}, extrapolated where necessary from the 
end of 2011 using the average indexation for 2011 (shown dashed). 
In order to provide a basis of comparison for the loans that is relatively
neutral to the underlying monetary expansion and its effect on the monetary
unit of measurement, the base amount of the loan 
in the figures is 80\% of the average price for a house in Reykjavik 
during the year the loan was taken out.\footnote{The source for house prices is 
{\th}j\'o{\dh}skr\'a \'Islands (Registers Iceland 
\url{http://www.skra.is/lisalib/getfile.aspx?itemid=7850})}. 
The monthly repayment for the mortgage is shown in red, and the outstanding
principal in black. 
\clearpage   
\subsection{Interaction with Monetary Regulation}
Increases in the consumer price inflation rate have two fundamental 
causes: either scarcity of the goods
and services in the index causing a rise in prices due to supply
and demand factors, or expansion in the money supply. Conversely excess
supply will cause a drop in prices, as will a reduction in the
quantity of money. This relationship
follows directly from the quantity theory of money\cite{fisher.1911}
which is one of the fundamental equations of Economics 
\footnote{$P \approx M/Q$ where P is the aggregate price level, 
M is the money supply and Q is the quantity of monetary
transactions performed with M. See Mallett\cite{mallett.2012.2} for comments
on the reason for the removal of V from this version of the equation}. 
Analysis of price data must consequently be performed with some caution
since for any given price movement there are two possible causes, which
without reference to the underlying money supply data it is impossible
to untangle. This result applies generally and includes relationships
between currencies, which are a balance of export and import demands
and the relative expansion of the money supplies of the countries
concerned. Iceland's money supply has usually expanded 
at considerably higher rates than those of its trading parties, 
consequently the foreign exchange 
fluctuations which are often cited as the direct cause of inflation in 
its banks' annual reports are better regarded as a consequence of  
money supply growth rather than its cause.
\par
Money supply expansion can similarly arise from two different sources.
Banks as part of their normal operations create money in the form of
a bank deposit, when a loan is made. This money is then destroyed
as loan principal is repaid. Whether or not there is a net increase
in the money supply depends on the
regulation of this process such that the rate of new lending 
is matched by the rate of repayment of the principal on existing loans 
across the entire banking system. Physical printing of money by
the government, especially in the gold standard banking regimes that 
were regulated by central bank reserves will also lead to rapid 
hyperinflation, since the additional physical money(which is treated
as an asset) can simultaneously trigger a multiplier
expansion in bank deposits(a liability) by effectively lifting the
reserve limits. In Basel regulatory frameworks this effect is still
present, but since lending is also constrained by the capital limits
its effects are not so dramatic.
\par
There appear to exist several problems in
the current regulatory framework which relies on a combination
of central bank reserve regulation, and controls on the ratio of bank 
capital to their lending portfolio (see Mallett for 
specific details\cite{mallett.2012.1}). 
Modern banking systems demonstrate a wide range of expansion rates
ranging from 1.27 times in the last decade in Germany, 2x in the USA,
up to Russia whose money supply has increased by 20 times since its
last revaluation in 2000.  Iceland's expansion
rate of 6.76x in the post bank privatisation period 1999-2008 is at the 
high end of the range.   
\par
There are then certainly grounds for suspicion that index-linked loans
may be causing systemic side effects, and in particular interacting with
the money supply.   Whether or not this is actually the case 
appears to depend critically on the precise bookkeeping
operations that are used by the banks to perform the indexation on the
loans they retain on their asset books.
\par
\subsection{Bookkeeping of Icelandic Indexed Linked Loans}
\input{loan1.tex}
Table \ref{tab:bank_init} shows a simplified view of the asset and liability
ledgers for a single bank, and their relationship with those of the central
bank. Under double entry bookkeeping, all operations on the ledgers must 
be performed as (credit, debit) tuples. Each single action, depositing
money for example, results in two matching transactions, as a result of 
which the total amount of assets must always be equal to the total amount
of liabilities. In this example we show an initial state with 
two deposit accounts, a matching quantity of loans, 
capital holdings of 1000, and central bank reserves (an asset) of 210
with a corresponding deposit (a liability from the central bank's
perspective) at the central bank. The cash 'asset' represents
either physical cash, or asset accounts at other banks. We will assume
that the risk weighted multiplier of 50\% for mortgages (Basel 2) 
applies, and so that bank is well within its capital limits for
its lending book.
\par
It should be mentioned that the reality of bank operations differs 
significantly from the description commonly found in Economic textbooks. 
Central bank reserves are not held back from customer's deposit accounts 
as is often implied, but are separate holdings, and are classified as a 
bank asset, and not as a liability as the customer's deposit accounts 
are.\footnote{Strictly, the central bank reserve requirement was
a fraction of the physical cash deposited at the bank. The customer's 
deposit account was the liability entry reflecting this deposit. 
Bank loan's were written as assets, with a matching deposit.
The introduction of cheques and other forms of direct transfer
between customer accounts, effectively introduced two forms 
of equivalent, but not identical money into the monetary system.}
Historically writers in the early
20th Century such as Keynes\cite{keynes.1929} would distinguish
customer deposits as 'deposit money' or similar, and there has been
considerable debate about its precise status. Today with electronic
transfers being used for almost all monetary transactions
in Iceland and elsewhere, it is the amount of money held in bank deposits
that is most critical for determining the price level, while
the money classified as assets, either in the form of physical 
cash or deposit accounts at other banks including the central 
bank, plays both a regulatory 
role, and is also critical in providing liquidity for inter-bank clearing
relationships.
\par
\input{loan2.tex}
To illustrate with an example, Table \ref{tab:loan_cust} shows 
a bank loan being created to a customer at Bank A, with the 
accompanying double entry bookkeeping operations. In this
example, Bank A makes a loan of 500 to its customer C1, the resulting
ledger changes are shown in blue. Besides the increase in its deposits
and loans books, it also increases its central bank reserve holdings 
to maintain its required regulatory position. As it was overcapitalised
before the loan was made, it remains within its 
Basel Capital limits. When the loan principal is repaid 
is is reduced, as is the deposit account of the depositor making 
the repayment. 
When interest is paid on the loan there are no direct money supply 
implications as is also sometimes believed: a debit is made to the 
customer's account, and a credit to the bank's interest income account. 
Since both of these are classified as liabilities there is no net change 
to that side of the balance sheet, and the transaction is money supply neutral. 
\par
Under double entry bookkeeping rules any change to a single
ledger, such as loan assets, must be matched by a corresponding
debit or credit to another ledger, maintaining the fundamental accounting
equation $Assets = Equity + Liabilities$. Consequently when loan 
indexation causes the book value of a bank loan asset to increase, there must
be some form of matching operation to maintain the balance across the
bank's books, and it is in the handling of this event where the 
possibility of potential interactions with the money supply arises.
The impact of the index-linked loans on the banking system
depends fundamentally on this rather obscure bookkeeping question - 
what is the compensating operation for the increase in the loan's 
principal(an asset), caused by the combination of indexation and 
the negative amortisation structure of these loans?  There appear 
to be two alternatives.
\par
One approach would be to use a contra-asset account. Contra-asset
accounts are asset accounts with a credit balance, which effectively
act as offsetting (negative) balances on assets.\footnote{In double
entry bookkeeping the arithmetic operation of 'credit' and 'debit' 
within a balanced ledger system depend on the
side of the ledger they are being applied to, rather than their english
meanings. Credits to an asset account will reduce the balance for example, 
while debits increase it, and the opposite applies to liability
accounts.  The general rule is that all operations must consist 
of (credit, debit) tuples, and the individual Asset/Liability 
classification of ledgers is fundamentally determined by the need to 
ensure equality between 
the total of assets and liability ledgers as these operations are performed.
In bank accounting the classification 
of asset and liability accounts is sometimes the opposite to that used
by non-banks: for example bank deposits are classified as liabilities, and 
loans to customers as assets.} This would maintain the balance sheet
identities and would not have any direct monetary impacts. 
\par
There is some evidence from
B\'una{\dh}arbanki \'Islands' Annual 
Report in the early 1980's that suggests this approach may have been 
used earlier. Between 1981-1994 they show an Asset balance for accrued interest 
and indexation, (\'Afallnir vextir og ver{\dh}b{\ae}tur).
There is however no obvious
correspondence between the growth in this item and the entry for
index-linked loans. Between 1984-5 for example, outstanding principal
for index-linked loans grew from 752 million ISK to 1,300 million ISK,
whilst the growth in the accrued interest and indexation account was only
102 million ISK, and the annual inflation rate was 
34\%.\footnote{B\'una{\dh}aribanki was one of the 
three main banks in Iceland up until its merger with Kaupthing in 2003. 
Kaupthing was nationalised following the 2007 crash, and is now known as
Arion Bank.}
\par
\begin{figure}[ht]
\begin{center}
\includegraphics[width=10cm, clip]{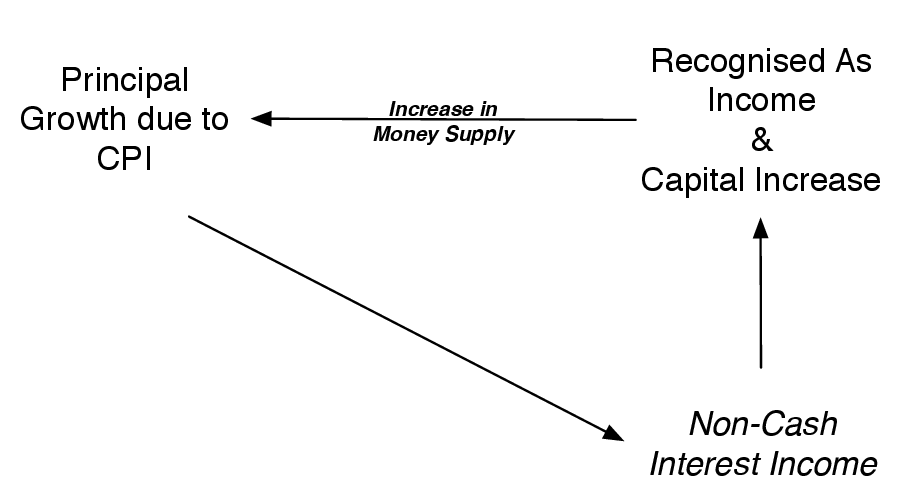}
\caption{Indexation -> CPI Feedback Loop}
\label{fig:vl_feedback}
\end{center}
\end{figure}

The bookkeeping which appears to have been used for these loans 
is shown in Tables \ref{tab:principal_growth1} and 
\ref{tab:principal_growth2}.\footnote{This description follows what was 
reported to us in discussions with Icelandic Auditors specialised in banking.} 
The matching balance for any increase in the loan 
principal due to indexation is first credited 
to the liability non-cash, interest income received account. 
At the end of the year, 
banks are allowed to recognise this account as income, subject to 
regulatory controls on loss provisions and allowances for loan defaults,
and regulatory capital. When this occurs the money credited
to this account as a result of loan indexation is recognised as 
an expense or dividend payment and transferred to a deposit account.
At this point it becomes money and can affect the price level. The
result is a feedback loop operating on the money supply, as shown 
in Figure \ref{fig:vl_feedback}. 
\par
The following exert from Chapter VI of the Rules on the financial 
statements of credit institutions No. 34  of 3 November
2003\footnote{\url{http://en.fme.is/media/utgefid-efni/rules_834_2003.pdf}
(page 14)}, also supports this treatment, where it details the reporting 
requirements for indexation. In particular:
\begin{quote}
Profit and loss account item 1.2, "Interest on loans and advances etc.", 
shall comprise interest receivable, \emph{indexation} and commissions 
receivable, calculated on a time basis or by reference to the amount of the 
claim, on Assets item 4, "Loans and advances", including 
credited discount on loans and advances
\end{quote}
\par
To illustrate these operations in more detail, 
Table \ref{tab:principal_growth1} shows the
result of an increase of 500 in the loan principal at Bank A and the 
corresponding
entry in the non-cash income liability ledger. Bank B is set at the initial
condition for comparison.
Table \ref{tab:principal_growth2} then shows the subsequent recognition 
of part of the amount as income
and payment to the deposit holder A.C1 as a bank expense as salary, dividend
payment to shareholder, payment for services, etc. A small amount would
also need to be withheld to cover the increased Basel capital reserve required 
for the new loan value, which is not shown in this example.
\input{vl_iceland}
\par
It is not known why the loans were structured to negatively amortise the
indexation component, since it is not strictly necessary. The negative
amortisation is created by the application of an annuity factor to 
the repayment schedule which decays over time, creating progressively
higher capital repayments. The point at which the growth in the 
repayment amount overcomes the negative amortisation also depends 
on the underlying growth in the principal - as can be seen when
comparing the hyperinflationary period with later loans.
The negative amortisation of the loans does make them initially
significantly less expensive than the alternative, and this may 
have been required to make them appear affordable.  It is also not
known if this bookkeeping treatment is unique to Iceland. Other
countries, notably Chile have attempted to use indexation to control
high inflation (see Lefort\cite{lefort.2002} for a survey of the
literature on Chile's experience); negatively amortized loans
are also available in other countries, and played a part
in the US subprime collapse. 
\subsection{Quantitative Analysis}
Analysis of how much additional inflation these loans cause is difficult.
As long as there is sufficient loan demand, the current banking system 
will expand the money supply to whatever limit is otherwise imposed by 
the regulatory framework it is operating under. Since the Basel framework 
imposes no absolute limits on total capital levels across the banking 
system, and central bank reserve requirements are not applied to all
accounts, the expansion rate seems to primarily depend on the rate of 
increase of the capital reserve, which in turn depends directly on the 
profitability of the banks, and their
individual lending decisions. A bank could choose not
to expand their capital reserve if they felt that the new lending being
requested was too risky for example, although this doesn't appear to 
have been a factor in Icelandic banking decisions during
the bubble period. The 
question is how much of the 
capital expansion can be attributed to these loans, and how much would
have occurred anyway?  While we cannot put an absolute figure 
on this, we can estimate upper and lower bounds.
\par
For a very brief period between July 1993 and April 1995, the annual
inflation rate in Iceland fell below 2\%. This is
under the threshold for negative amortisation to occur on these loans,
and so during this time there would have been little or no
monetary growth due to this factor. The effect of this can be seen in
Figure \ref{fig:vl_actual_25}c, where principal growth flatlines
for 2 years, and then subsequently increases. 
\par
The annual M3 series from Se{\dh}labanki \'Island's shows
a growth in M3 of 2.3\% for 1994-95, so we can regard this as an 
approximate lower bound on the expansion rate of the Icelandic 
banking system without the influence of the growth in principal 
caused by indexation. 
\begin{figure}[ht]
\begin{center}
\includegraphics[width=10cm, clip]{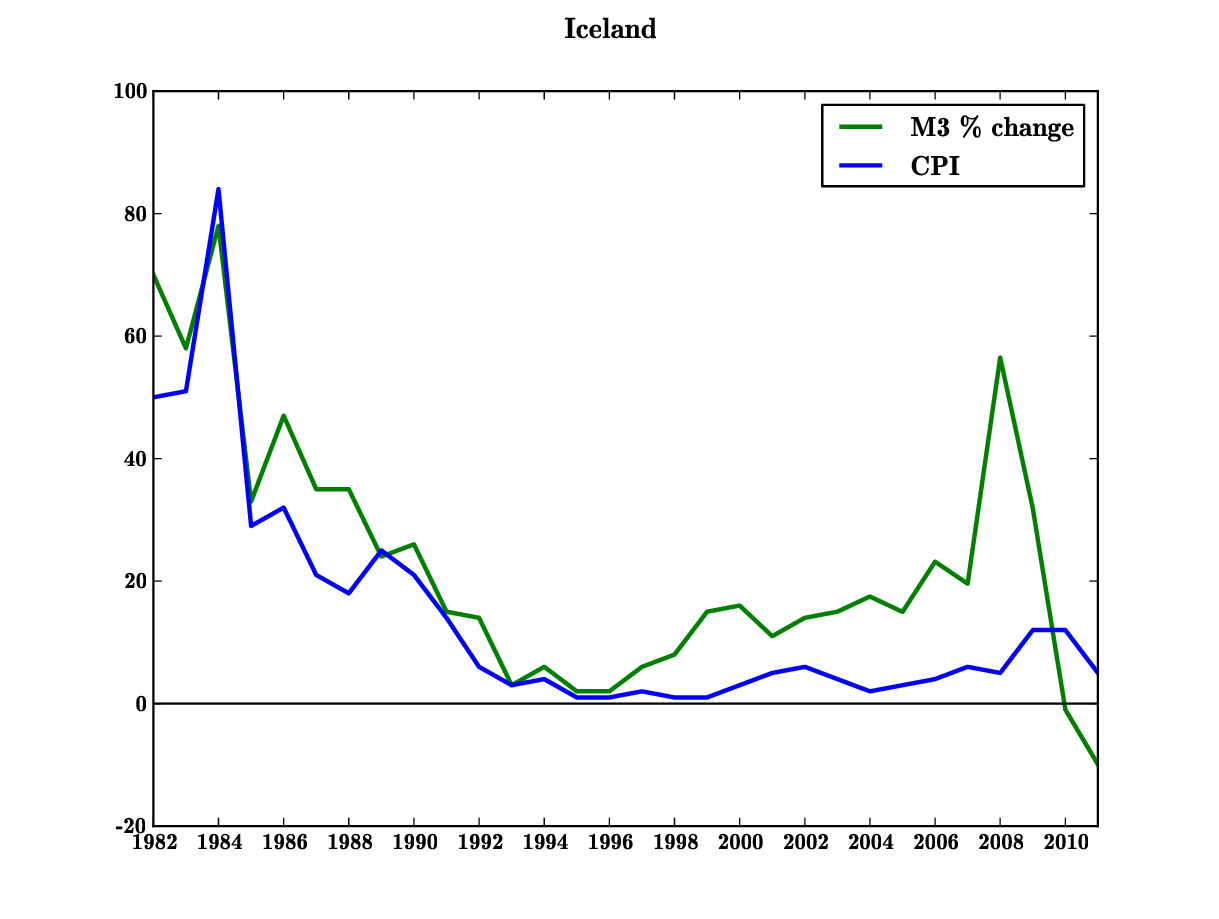}
\caption{Iceland CPI vs M3 \% change 1982-2011}
\label{fig:iceland_m3_cpi}
\end{center}
\end{figure}
Figure \ref{fig:iceland_m3_cpi} shows the relative growth of the Icelandic
CPI versus the M3 growth over the 1980-2012 period. If we exclude the
hyperinflationary periods between 1967-1990, and 
2005-8\footnote{This latter period featured the main Icelandic
Banks developing several innovative approaches to financing capital
expansion, which are currently the subject of legal action.}
we can see
that that the M3 expansion rate increased from 6.86\% in 1996 to
14.97\% in 2004. So we can place the upper bound on monetary
expansion due to these loans between 4.56\% and 12.67\%, bearing
in mind that the precise amount of negative amortisation with 
the loans is a function of the inflation rate itself. Consequently
there is 
a positive feedback loop involved, illustrated in 
Figure \ref{fig:vl_feedback}.
which implies that the rate of expansion is likely to increase over
time. 
\par
To compare the additional cost of these loans 
Table \ref{tab:repayment} shows 25 and 40 year repayment totals for 
Icelandic versus fixed rate US loans at a range of interest rates. 
\begin{table}[ht]
\centering
\begin{tabular}{lccccc}
Amount     & Base Rate & Indexed (1990) & Fixed Rate    & Indexed (1990) & Fixed Rate\\
           &           & \multicolumn{2}{c}{25 Years} & \multicolumn{2}{c}{40 Years} \\
\hline
20,000,000 &    4.0\%  &    55,981,097   & 31,670,200 & 134,458,782 & 40,122,000  \\  
20,000,000 &    5.0\%  &    62,000,204   & 35,075,400 & 155,131,850 & 46,290,800  \\  
20,000,000 &    7.0\%  &    74,959,290   & 42,406,700 & 199,926,304 & 59,657,400  \\
\end{tabular}
\caption{Total cost of 40 year Indexed Linked and Fixed Rate loans}
\label{tab:repayment}
\end{table}
Besides the extra cost of the loans, Table \ref{tab:repayment} also 
illustrates a less obvious aspect of the loans, their profile depends 
critically on when they were taken out. While it might be expected that 
the loans would behave similarly whenever they were created, this is 
not the case. Borrowers in the 1980's
experienced hyperinflationary increases in their loan principal, but 
their salaries were also indexed. In the
1990's, a period of relatively low monetary expansion occurred, at the same time
as productivity greatly increased, with the result that inflation was low,
and the negative amortisation of the loans was minimised. In the 2000's however,
borrower's experienced both high inflation and higher interest rates
for the fixed interest rates on the loans. While salaries generally rose 
above the inflation
rate during the early part of the decade, post-2008 salaries generally 
dropped significantly.  This also helps to explain the disproportionate 
difference in the cost of the 25 versus the 40 year index-linked loans,
once the negative amortisation component is taken into account.
\section{Conclusion}
One of the striking problems of current banking system research is that
there is relatively little focus on an integrated view of the entire
system's behaviour. Consequently when reforms are proposed they typically 
address single issues affecting one part of the system, and not 
infrequently - as the index-linked loans illustrate only too well - create new 
problems elsewhere. This paper should of course also be 
read in that context. A far deeper understanding of the larger implications
for monetary systems of indexation instruments is required than is 
currently available, and this is an area requiring considerably more
research. 
\par
This failure to identify an integrated understanding of banking system
behaviour has long proved problematic for policy makers.
The indexed-linked loans were originally introduced with the objective
of stabilising the Icelandic banking system during a period of hyperinflation
triggered by government money printing.
That government printing of money can cause hyperinflation is of 
course well known, but that one of the causes of this form of 
runaway monetary expansion arises from interaction with the banking system's
regulatory controls is not. That indexing loans would cause monetary
expansion due to obscure details of the bookkeeping treatment used, is
also not the most obvious of side effects.
\par
There are other objections that can be raised against the index-linked
loans besides their inflationary influence though.  The formula used for their 
calculation are not readily available; their historical calculation also rests
on time series data on the applied indexation rate that is difficult 
to locate; and it is has consequently proved extremely difficult for 
borrowers to validate their payment schedule. Since changes are 
periodically made to how the indexation
is calculated their future behaviour is also impossible to predict,
considerably complicating individual financial planning.
With the newer form of the loan that has been issued since 2005, the base
interest rate can also be arbitrarily adjusted
by the lender after an initial period, and there appear to be
no contractual limits on this. 
\par
The negative amortisation of the principal, which is responsible
for the loan's interaction with the CPI, also violates normal principles
of prudential borrowing. Borrowers do not typically begin
repaying capital until 15-20 years into the loan, and they incur extraordinary
debt loads as a direct consequence, especially with the 40 year loans. 
Unfortunately these loans
have been the predominant form of lending in Iceland for over 30 years, 
and the long term outlook is not known: in particular excepting 
loans which are repaid through property sale, what percentage of these loans
are ever successfully repaid?
\par
The long term prognosis for recent borrowers in particular is grim.  
As shown in Figure \ref{fig:vl_actual_40}, the
combination of high inflation before the crash, and the high base interest rates
at the beginning of the 21st century will make it extremely hard if 
not impossible for many borrowers to fully repay these loans, and while
statistics on the proportion of 40 year loans are not
available, they are believed to be the majority of these loans. 
Icelandic policy since the crash has been to attempt to protect vulnerable
and lower income groups, and
Stef\'{a}n \'{O}lafsson\cite{olafsson.2011} has presented 
a detailed analysis of household situation and policy responses showing
the partial success of this. 
However, the various forms of relief offered have often penalised
those who acted responsibly during the bubble - in 2008 for
example borrowers were allowed to apply for a revaluation
of their loans to no more than 110\% of the value of the house, effectively
penalising those who had substantial personal equity in their 
houses. Additional relief has also
been provided through tax deductible interest rate subsidies to a maximum 
of 900,000 ISK for married couples in 2012. As a consequence house
price levels have been largely maintained, but there has been considerable
pressure on rental levels, leading to high rents, and then to compensating 
increases in tax funded rental subsidies.
\par
\begin{figure}[ht]
\centering
\begin{tabular}{cc}
\subfloat[2000 11,333,000 ISK @ 7.8\% base rate]
   {\includegraphics[width=7.5cm]{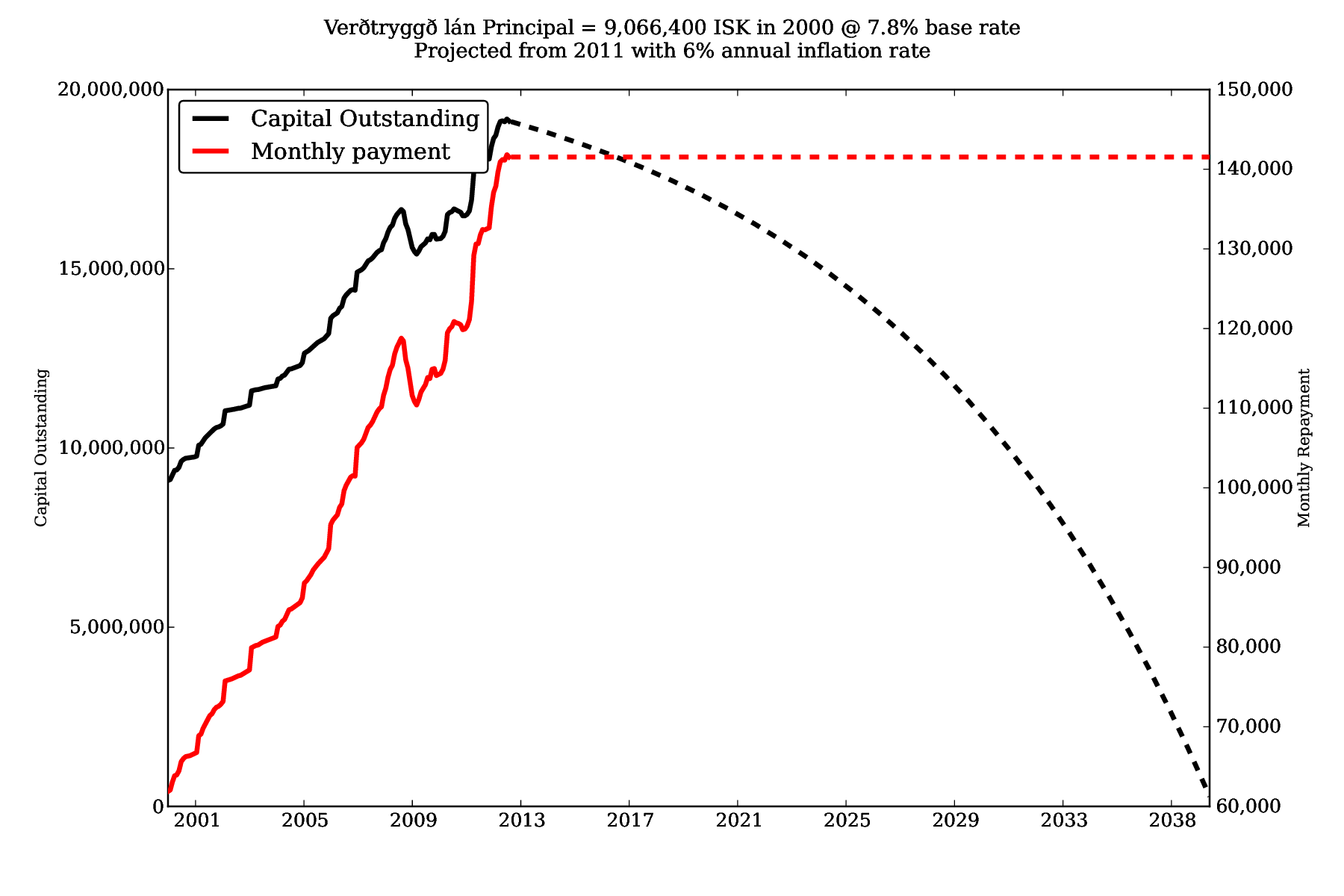}}   &
\subfloat[2005 20,779,000 ISK @ 4.2\% base rate]
   {\includegraphics[width=7.5cm]{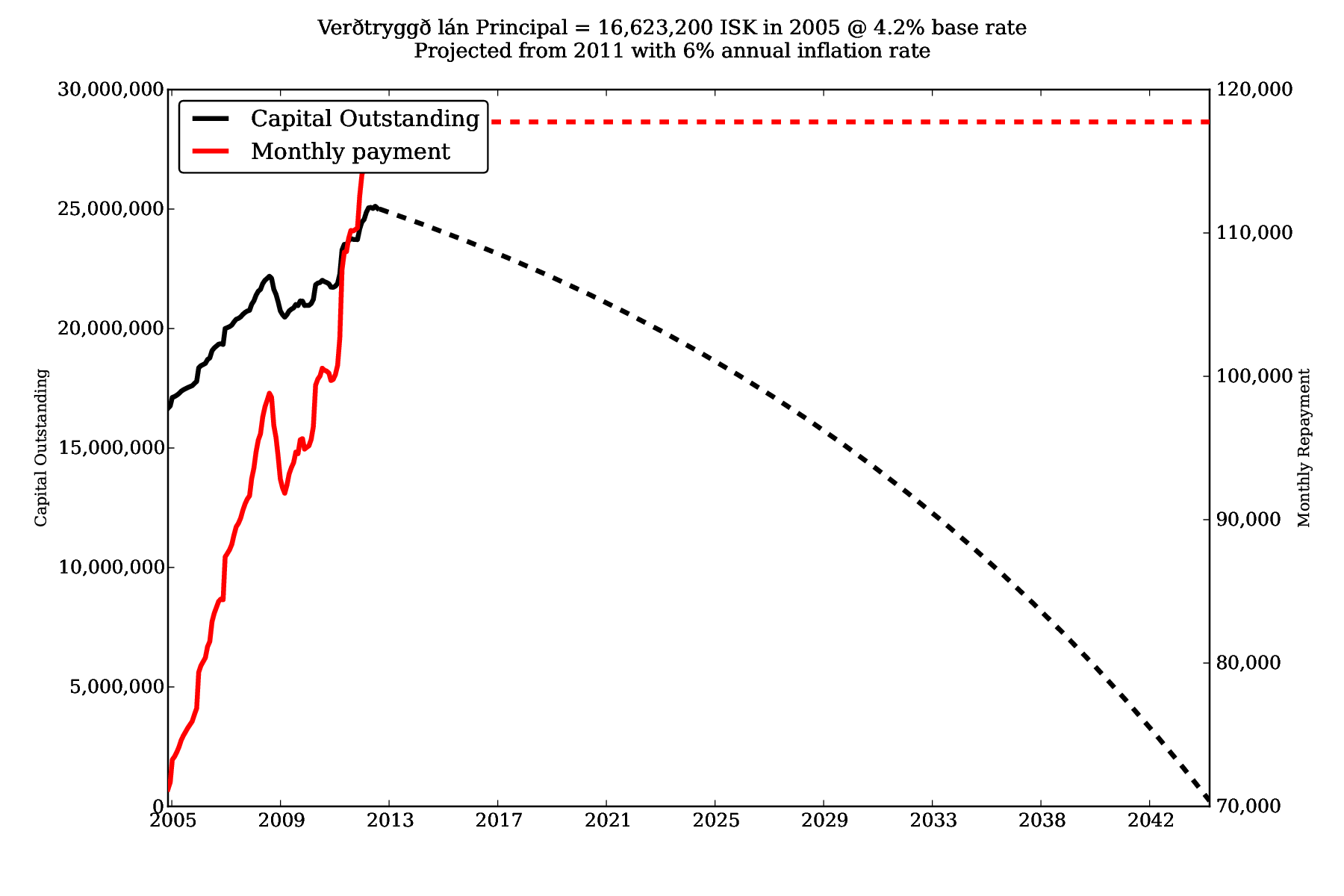}}   \\
\end{tabular}
\caption{Projected repayment on 40 year loans with 0\% inflation.}
\label{fig:vl_projected_40_0}
\end{figure}
\par
A further complication lies in the ownership of the loans.
Indexed linked loans were available from the Icelandic Banks, and
from the Housing Finance Fund(HFF)\cite{sveinsson.2004} which is a government
owned provider of mortgage credit. They were also securitized. A 2006 Report by 
Kaupthing Bank\cite{kaupthing.2006} showed ownership split between the 
Icelandic Banks(20\%), Mutual Funds(9\%), 
private holdings(8\%) foreign investors(46\%) and 
the Icelandic Pension Funds(17\%). While within Iceland there
are widespread calls for indexation to be removed from these loans,
ownership by the pension funds who are required to index link their
pension payments, and in particular the high foreign ownership 
considerably complicates any formal adjustments.
\par
It should be emphasised that only index-linked loans directly owned by 
commercial banks or similar institutions performing fractional
reserve lending can contribute to money supply growth through the 
mechanisms described in this paper. However all borrowers with
these loans are affected by the consequent growth in the money supply.
Since substantial quantities of these loans were securitized and sold
abroad (the Glacier Bonds), all Icelanders are potentially affected
by the resulting foreign exchange imbalances which are being temporarily
managed through capital controls. It seems entirely too probable that
this will cause further feedback issues when the capital controls
are lifted. If loan indexation causes higher
inflation, then currency outflows on these loans will increase,
weakening the Krona exchange rate, and further increasing inflation
as the cost of imported goods rise - which then increases monetary
expansion through the indexed loans, leading inexorably to higher
inflation.
\par
In addition borrowers are
now faced with loan and money supply growth in the banking system 
originating from the newly introduced fixed rate loans which are not
negatively amortized.\footnote{Since the crash
loans with a fixed interest rate for the first 5 years, and varying subsequently
have been introduced by the banks.} While the Icelandic 
money supply contracted
by 10\% in the immediate aftermath of the crash, it grew by 7\% in 2011.
New borrowers with these loans will benefit from the monetary expansion
caused by the indexed-linked loans, as the resulting inflation reduces
their debt burden; just as it increases the debt burden for those
with indexed-linked loans.
Consequently it seems highly probable that over time those with indexed-linked 
loans who are unable to refinance into fixed rate loans, 
will be effectively trapped into eventual bankruptcy, as
a rapid divergence in individual situations divides house owners with
growing debt from indexed-linked loans, from those with declining debt
with the newer fixed rate loans.
\par
Somewhat ironically, given the original reasons for introducing the
loans was monetary stability, there is a possible solution that would 
effectively remove
indexation from these loans, without the need for any formal
loan modifications. This is quite
simply to stabilise the money supply to 0\% growth, which could then
be expected to similarly stabilise the CPI. 
As shown in Figure \ref{fig:vl_projected_40_0}, when the CPI is 3\% or less,
these loans behave identically to compound interest fixed rate loans, and the
negative amortisation and consequent money supply interaction ceases. 
\par
This would also help to mitigate the pension fund exposure to these loans,
as in a constant money economy with low or negligible 
inflation their indexation 
requirement towards their pensioners is also resolved. They would 
also benefit from a considerably lower 
default rate on their loans than would otherwise be the case, resulting
in lower capital losses.
An economy with a constant money supply would also
provide considerably clearer signals to its policy makers through the 
price mechanisms, since
the distortions caused by varying rates of monetary expansion would be
removed, and this would considerably aid the establishment of sound
and prudent policy direction. 
\par
Achieving this goal would require adjustments to
the regulatory banking framework, and a legal mandate established to 
support it. Mechanically speaking there are several ways the problem 
could be approached.  The simplest would be to adjust the current 
Basel Regulatory framework to maintain
a constant capital amount across the banking system, while also creating
a unified risk weighting for all types of lending, and in particular removing
the 0\% weighting on lending to governments. 
\par
The larger economic effects of monetary stability are not known, since
this does not appear to have ever been achieved during the fractional 
reserve banking era. We suspect that one side effect would be that many 
of the economic indicators that we are used to, steadily increasing prices, 
steadily increasing government debt, would look quite different with a 
constant money supply, since
the distortion introduced by the change in the unit of measurement
for these statistics - money - would be removed.
However not enough is known about 
any long term side-effects this might create for liquidity provisions 
within the banking system as money is transferred between banks. Central
bank inflation targets (typically 2\%) arise from the observation that
low but positive rates of inflation appear to be beneficial - but the mechanisms
causing this effect are unknown, as is their general applicability. 
There would also be potential competitive issues between banks, if 
mechanisms were not developed for banks to trade capital within the system,
but potential stability issues if they were. However it can be
argued that these problems are also present in the current system.
A medium term policy of stabilisation using this approach, careful
monitoring, and an accelerated
research program into longer term regulatory frameworks capable of providing
monetary stability, in conjunction with a deeper investigation into
the macro-economic consequences of these loans would seem to be the
safest approach, given the current state of economic knowledge, 
to deal with these issues.
\par
More problematic are the potential exchange repercussions if Iceland did
achieve monetary stability, which is the danger that improbable as it
might sound, the Krona
becomes a store of value. The general failure of modern banking regulation
to provide monetary stability has created a world of widely varying monetary
expansion rates, and Iceland is not in a position to support
significant investment demand for its currency. Until these issues
are addressed on a worldwide basis, small countries with relatively 
stable currencies are placed in difficult situations - as the recent 
experience of Switzerland shows.
This would present an interesting challenge
to Iceland's monetary authorities, and restrictions on Krona ownership, 
or other mechanisms might need to be introduced to protect the Icelandic
economy if this eventuality arose.
\par
There are other approaches that have been applied by countries with excessive
mortgage debt, for example the USA's deliberate expansion of its money
supply over the last 4 years. This approach is reliant both on the 
preponderance of fixed interest rate lending in its monetary system, and 
the unique position the
US dollar currently occupies as a global trading currency. As a solution 
it is not available to Iceland
simply because of the pre-dominance of index-linked loans in the monetary
system. 
\par
It is unlikely that any single solution, even quantitative stability,
is likely to be the only required measure in Iceland's situation. The
issues created by the indexed-linked loans are complex, and potentially
very far reaching, given their 40 year duration.
Monetary policies must be developed that are based on a
grounded understanding of the direct causes of inflation, rather
than simplistic attempts to address some of its symptoms 
within a complex and poorly understand financial environment.
A co-ordinated systems based approach is required, and one that is tailored
to the unique circumstances that Iceland, and its citizens finds itself
in, rather than reliant on economic policy originating from very different 
monetary systems. 
\section*{Acknowledgements}
I would like to thank Gu{\dh}mundur \'{A}sgeirsson and Andri M\'{a}r \'{O}lafsson
for their invaluable help with background research and Icelandic
sources; Valborg Stef\'ansd\'ottir and Anton Holt at the Library of the
Central Bank of Iceland for their assistance with access to historical 
information on the Icelandic Banking System;
Dr. J\'on {\TH}\'{o}r Sturluson for discussions on fractional
reserve banking mechanisms, and with Einar J\'{o}n Erlingsson 
for assistance with the formula and calculations for the loans; 
Charles Keen, David Gudjonsson for their 
considerable help with review comments; and Dr. Kristinn Thorisson for
his extraordinary support and guidance with the fundamental research
on banking systems this paper builds on.
The author takes sole responsibility for any remaining errors.
\include{biblio}

\input{AppendixA}
\end{document}

%% file: titlepage_van.tex
\title{An examination of the effect on the Icelandic Banking System of
Ver{\dh}trygg{\dh} L\'{a}n (Indexed-Linked Loans).}
\author{Jacky Mallett  \\jacky@ru.is}
\date{February 15th, 2013} 
\maketitle

%% file: loan1.tex
\begin{table}[ht]
\centering
\caption{Simplified Bank Ledger Example}
\begin{tabular}{r | l c l r|r r}
\multicolumn{2}{c}{Central Bank}      &     &            & \multicolumn{2}{c}{Bank A}&          \\
                 Assets & Liabilities &     &            & Assets      & Liabilities &          \\
\cline{1-2} \cline{5-6} 
                 &             &     & Loans      & 10000  & 5000      & Deposit A.C1    \\
                 &             &     &            &        & 5000  	   & Deposit A.C2 \\
                 & 210         &     & Reserves   & 210    &   	       &                          \\
       210       &             &     & Cash       & 790    &   1000      	& Capital  \\
\cline{1-2} \cline{5-6}
     210         &     210     &     & Total      &   11000      	&  11000       	&          \\
\end{tabular}
\label{tab:bank_init}
\end{table}

%% file: loan2.tex
\begin{table}[ht]
\centering
\caption{Loan to Bank's own customer}
\begin{tabular}{r | l c l r|r r}
\multicolumn{2}{c}{Central Bank}      &     &            & \multicolumn{2}{c}{Bank A}&          \\
                 Assets & Liabilities &     &            & Assets      & Liabilities &          \\
\cline{1-2} \cline{5-6}
                 &             &     & Loans      &\color{blue}{10500}  &\color{blue}{5500}      	& Deposit A.C1 \\
                 &             &     &            &           		   &  5000      	& Deposit A.C2 \\
                 &\color{blue}{210}   & & Reserves&\color{blue}{210}   &           	    &          \\
\color{blue}{210}&             &     & Cash       &\color{blue}{790}   &   1000      	& Capital  \\
\cline{1-2}\cline{5-6}
     210         &     210     &     & Total      &   11500      	&  11500       	&          \\
\end{tabular}
\label{tab:loan_cust}
\end{table}
\par

%% file: vl_iceland.tex
\begin{table}[ht]
\centering
\caption{Principal Growth of 500 and Credit to Non-Cash Income}
\begin{tabular}{r | l c l r|r r}
\multicolumn{2}{c}{Central Bank}      &     &            & \multicolumn{2}{c}{Bank A}&         	\\
                 Assets & Liabilities &     &            & Assets      & Liabilities &         	\\
\cline{1-2} \cline{5-6}
                 &             &     & Loans      &\color{blue}{10500} 	&   5000 	     	& Deposit A.C1   	\\
                 &             &     &            &           	        &   5000      			& Deposit A.C2   	\\
                 &             &     &            &           	        &\color{blue}{500}     	& Non-cash Income(Bank)\\
                 &   20        &     & Reserves   &   20      	        &           			&            	\\
            40   &             &     & Cash \& Eq &   980      	        &\color{blue}{1000}    	& Capital    	\\
\cline{5-6}
                 &             &     & Total      &\color{blue}{11500}     	&\color{blue}{11500}      	&            	\\
\\
\\
                 &             &     &            & \multicolumn{2}{c}{Bank B}&            	\\
\cline{5-6}
                 &             &     & Loans      &   10000     &   5000      	& Deposit B.C3   	\\
                 &             &     &            &           	&   5000      	& Deposit B.C4   	\\
                 &  20         &     & Reserves   &   20      	&           	&                	\\
                 &             &     & Cash \& Eq &   980      	&   1000      	& Capital        	\\
\cline{1-2}\cline{5-6}
          40     &  40         &     & Total      &   11000     &  11000       	&                	\\
\end{tabular}
\label{tab:principal_growth1}
\end{table}
\begin{table}[ht]
\centering
\caption{Recognition and expense payment}
\begin{tabular}{r | l c l r|r r}
\multicolumn{2}{c}{Central Bank}      &     &            & \multicolumn{2}{c}{Bank A}&         	\\
                 Assets & Liabilities &     &            & Assets      & Liabilities &         	\\
\cline{1-2} \cline{5-6}
                 &             &     & Loans      &   10500      	&   5000      	& Deposit A.C1   	\\
                 &             &     &            &           	    &\color{blue}{5450}     	& Deposit A.C2   	\\
                 &   20        &     & Reserves   &   20      	    &           	& Non-cash Income(Bank)\\
            40   &             &     & Cash \& Eq &   980      	    &\color{blue}{1050}     	& Capital    	\\
\cline{5-6}
                 &             &     & Total      &   11500      	&  11500       	&            	\\
\\
\\
                 &             &     &            & \multicolumn{2}{c}{Bank B}&            	\\
\cline{5-6}
                 &             &     & Loans      &   10000      	&   5000      	& Deposit B.C3   	\\
                 &             &     &            &           	&   5000      	& Deposit B.C4   	\\
                 &  20         &     & Reserves   &   20      	&           	&                	\\
                 &             &     & Cash \& Eq &   980      	&   1000      	& Capital        	\\
\cline{1-2}\cline{5-6}
          40     &  40         &     & Total      &   11000      	&  11000       	&                	\\
\end{tabular}
\label{tab:principal_growth2}
\end{table}

%% file: biblio.tex
\bibliography{finance}
\raggedright
\bibliographystyle{unsrt}

%% file: AppendixA.tex
\appendix
\section*{Appendix}
\section*{Calculation of Indexed Linked Loans}
There are two forms of Ver{\dh}trygg{\dh} lan with slightly different
repayment profiles, Fixed Amortization and Fixed payment. In both 
cases repayment is structured as an annuity, with the annuity factor
partially determining the capital repayment. As a consequence,
for rates of inflation above 2-3\% (depending on the duration of the loan),
repayments during the first years of the loan may not cover principal
repayment and the remained is negatively amortized. This leads to 
considerable variation in the point at which principal repayment does
begin, depending in part on the duration of the loan, and on the
prevailing rates of inflation during the loan.
\subsection*{Fixed Amortization(real terms)}
For an $N$ period loan with an initial principal of $X_0$ and a fixed
rate interest rate component of $r$. Inflation is measured as the
percentage change in the CPI, $\pi_t = \frac{CPI_t - CPI_{t-1}}{CPI_{t-1}}$.

\begin{flalign*}
\text{Amortization in real terms:}&&  A{}_t^r &= \frac{X_0}{N}&  && \\
\text{Amortization in nominal terms:}  
&& A{}_t^n  &= A{}_t^r\frac{CPI_t}{CPI_0}                        && \\
&&          &= \frac{X_0}{N}\frac{CPI_t}{CPI_0}                  && \\
&&          &= \frac{X_0}{N}\prod{}_{j=1}^t (1 - \pi{}_j)        && \\
\text{The per (end of) period Principal in real terms:}        
&& X{}_t^r &= X_0 - \frac{X_0}{N}t                               && \\ 
&&         &= X_0(1 - \frac{t}{N})                               && \\
\text{The per period Principal in nominal terms:}              
&& X{}_t^n &= X_0 (1 - \frac{t}{N})\prod{}_{j=1}^{t}(1 + \pi{}_j) && \\ 
\text{Interest payments in real terms (paid at end of period):}
&& I_t^r &= rX_{t-1}^r                                            && \\
\text{Interest payments in nominal terms (paid at end of period):}
&& I_t^n &= r(1 + \pi_t)X_{t-1}^n                                  && \\
\\
\text{Total payments per period:} 
&& P_t &= A_t^n + I_t^n                                          && \\
&&     &= \frac{X_0}{N}\prod\limits_{j=1}^t (1 + \pi_j) + r (1 + \pi_t)X_{t=1}^{n} && \\
&&     &= \frac{X_0}{N}\prod\limits_{j=1}^{t} (1 + \pi{}_j) + rX_t^n
\end{flalign*}
\subsection*{Fixed payment (annuity in real terms)}
With terms as above, the fixed payment in real tems is derived from the 
standard equation of annuities:
\begin{flalign*}
&& P^r &= \frac{r}{X_0}\left(\frac{1}{1 - \frac{1}{(1+r)^N}}\right)  \\
\\
\text{Payment in nominal terms:}                   
&& P_t^n &= P^r\left(\frac{CPI_t}{CPI_0}\right)         && \\
&&       &= P^r\prod{}_{j=1}^t\left(1 + \pi_j\right)     && \\
\end{flalign*}
The breakdown of interest and capital payments is then calculated
recursively:
\begin{flalign*}
\text{Nominal Interest Payment:}                        
&& I_t^n &= r\left(0 + \pi_t\right)X{}_{t-1}^n        &&\\
\text{Nominal amortization:}
&& A_t^n &= P_t^n - I_t^n                             &&\\
\text{Nominal principal:}
&& X_t^n &= \left(1 + \pi_t\right)X_{t-1}^n - A_t^n   &&\\
\end{flalign*}
As the CPI changes over time, the re-payment schedule is adjusted through  
re-calculation of the 'fixed' payment.